\documentclass{iopart}
\usepackage{epsfig}
\usepackage{color}
\usepackage{amssymb}
\usepackage{times}

\newcommand{\rubidium}{$\phantom{|}^{87}{\rm Rb}$}
\newcommand{\Label}[1]{\label{#1}}

\newcommand{\ket}[1]{|#1\rangle}

\newcommand{\av}[1]{\langle#1\rangle}
\renewcommand{\b}[1]{{\bf#1}}
\newcommand{\eq}[1]{\begin{eqnarray}#1\end{eqnarray}}
\def\DRAFT{
\renewcommand{\Label}[1]{\label{##1}
{\hbox to 0cm{\textcolor{green}{\hss\em ##1\quad}}}}
 }
\begin{document}
\bibliographystyle{prsty}
\title{Theory of the Ramsey spectroscopy and anomalous segregation in ultra-cold rubidium}
\author{A. S. Bradley and C. W. Gardiner}
\address{School of Chemical
and Physical Sciences, Victoria University, Wellington, New
Zealand}
\begin{abstract}
The recent anomalous segregation experiment \cite{Lewandowski}
shows dramatic, rapid internal state segregation for two hyperfine
levels of \rubidium. We simulate an effective one dimensional
model of the system for experimental parameters and find
reasonable agreement with the data. The Ramsey frequency is found
to be insensitive to the decoherence of the superposition, and is
only equivalent to the interaction energy shift for a pure
superposition. A Quantum Boltzmann equation describing collisions
is derived using Quantum Kinetic Theory, taking into account the
different scattering lengths of the internal states. As spin-wave
experiments are likely to be attempted at lower temperatures we
examine the effect of degeneracy on decoherence by considering the
recent experiment \cite{Lewandowski} where degeneracy is around
$10\%$. We also find that the segregation effect is only possible
when transport terms are included in the equations of motion, and
that the interactions only directly alter the momentum
distributions of the states. The segregation or spin wave effect
is thus entirely due to coherent atomic motion as foreseen in
\cite{Lewandowski}.

\end{abstract}
\section{Introduction}
A topic of great interest in many particle quantum mechanics is
the effect of coherence in mesoscopic systems. The recent
experiment at JILA \cite{Lewandowski} shows transient highly
non-classical behaviour, characteristic of such phenomena. An
ultra-cold non-condensed gas of \rubidium\;  is harmonically
trapped and prepared in a superposition of two hyperfine states by
a two photon $\pi/2$ pulse. The superposition is allowed to evolve
for a range of different times and the densities of the two levels
measured across the trap; the data are then collected via
absorption imaging. In the cold collision regime of the experiment
the two states interact via slightly different S-wave scattering
lengths; moreover, the states experience different Zeeman shifts
in the presence of a magnetic bias field, and the combination of
these effects produces an effective potential between the two
states \cite{Lewandowski}. This {\it differential potential} was
characterized using the Ramsey spectroscopy technique, and the
segregation effect was explored for different potentials and atom
densities. When the differential potential is constant across the
cloud the motion is not observed. When the differential potential
has a gradient across the cloud the atoms appear to redistribute
in the trap, as long as the two states are in a partial
superposition. The dissipation of the coherent relationship
between the states is responsible for the transience of the spin
waves.

The original experiment \cite{Lewandowski} stimulated much work on
this system. Three theoretical treatments
\cite{Fuchs,Oktel,Williams2002} found good agreement between
simulations of the segregation dynamics and the evolution of the
experimentally measured distributions reported in
\cite{Lewandowski}. These works all use the spin operator notation
and treat the system as essentially a spin wave problem. More
recently, detailed imaging of the Bloch vector was carried out
during spin wave motion in \cite{McGuirk02}. Simulations of
linearized spin kinetic equations also showed good agreement for
the frequencies and damping damping times of linearized collective
modes \cite{Nikuni02}. We do not use a spin operator formalism in
this paper, although this has no effect on our results.

The outline of this paper is as follows. The equations of motion
are found from the Hamiltonian for the system in Section
\ref{sec:tfw}. The spectroscopy of the initial state is outlined
in Section \ref{sec:spec}. The interpretation of the Ramsey
experiment and the interaction energy are discussed in Section
\ref{sec:Ramsey}. In Section \ref{sec:cdamp} the issue of damping
is addressed. A Quantum Boltzmann equation describing the effects
of collisions on the distributions for states with different
scattering lengths is derived, and we discuss the relaxation time
approximations used in the simulations presented in Section
\ref{sec:sim}. In Section \ref{sec:spin} we consider the question
posed in the experimental report regarding the physical state of
motion of the atoms during the segregation.
\section{Theoretical Framework}\label{sec:tfw}
\subsection{Second Quantized Hamiltonian} The second
quantized Hamiltonian for a coherently coupled two state gas is
\begin{eqnarray}\Label{Hamil}
H &=& \int d^3{\bf x}\, \Bigg\{ \psi^\dagger_1 H_1({\bf x})\psi_1
+ \psi^\dagger_2H_1({\bf x})\psi_2
\nonumber \\
&& + {u_{11}\over 2}\psi^\dagger_1\psi^\dagger_1\psi_1\psi_1
 + {u_{12}}\psi^\dagger_1\psi^\dagger_2\psi_1\psi_2
 + {u_{22}\over 2}\psi^\dagger_2\psi^\dagger_2\psi_2\psi_2
\nonumber \\
&& + gE({\bf x})\bigg( \psi^\dagger_1\psi_2 e^{i\Omega t} +
\psi^\dagger_2\psi_1 e^{-i\Omega t} \bigg)\Bigg\},
\end{eqnarray}
where
\begin{eqnarray}
 H_{j}({\bf x}) =  -{\hbar^2\nabla^2\over 2m}+V_j({\bf x}) + \hbar\omega_j({\bf x})
\end{eqnarray}
for $j=1,2$, and $g$ is only nonzero during the initial two photon
pulse, creating a coherent superposition of the two hyperfine
states. Note that the Zeeman splitting of the transition frequency
is position dependent but this is absorbed into the effective
external potentials. The Heisenberg equations of motion for the
field operators are
\begin{eqnarray}\Label{Heom1}
i\hbar \dot{\psi}_1 &=&H_1({\bf x})\psi_1+u_{11}\psi^{\dag}_1
\psi_1\psi_1 + u_{12}\psi^\dag_2\psi_2\psi_1 +gE({\bf x})\psi_2
e^{i\Omega t},\\\Label{Heom2} i\hbar \dot{\psi}_2 &=&H_2({\bf
x})\psi_2+u_{22}\psi_2^\dag\psi_2\psi_2 +
u_{12}\psi^\dag_1\psi_1\psi_2+ gE({\bf x})\psi_1 e^{-i\Omega t}.
\end{eqnarray}

\subsection{Experimental Details}
 The experiment uses an ultra cold non-condensed \rubidium\;  gas held
in a harmonic trap at temperatures T$\sim 850\;$nK. The trap is a
cigar shape with frequencies $\omega_r = 2\pi\times 7\;$Hz and
$\omega_z = 2\pi\times 230\;$Hz. A $\pi/2$ two photon pulse is
used to create a superposition of the $|1\rangle\equiv |F=1,m_F =
-1\rangle $ and $|2\rangle\equiv |F=2,m_F = 1\rangle$ hyperfine
states from a thermal equilibrium $|1\rangle$ gas. The s-wave
scattering lengths between the hyperfine states $|i\rangle$ and
$|j\rangle$, $a_{ij}$, are $a_{11}=100.9a_0$, $a_{12}=98.2a_0$ and
$a_{22}=95.6a_0$, where $a_0$ is the Bohr
radius\cite{Lewandowski}. In the usual cold collision
pseudo-potential approximation this leads to the interaction
strengths
\begin{eqnarray}
u_{ij}=4\pi\hbar^2a_{ij}/m_R.
\end{eqnarray}
where $m$ is the \rubidium\;  mass. Since the scattering lengths
are so similar, and almost evenly split about $a_{12}$ we define
the interaction splitting
\begin{eqnarray}
\delta u&\equiv&u_{11}-u_{22}=5.3{4\pi\hbar^2a_0\over m},
\end{eqnarray}
and use the approximations
\begin{eqnarray}\Label{intapp1}
u_{11}-u_{12}&=&2.7{4\pi\hbar^2a_0\over m} \simeq \delta
u/2,\\\Label{intapp2} u_{12}-u_{22}&=&2.6{4\pi\hbar^2a_0\over m}
\simeq \delta u/2,
\end{eqnarray}
in the remainder of this paper.

 A magnetic bias field is used to change
the effective differential force exerted on the two states, and
can be chosen to either cancel or enhance the mean field force.
Segregation of the two species is observed via a subsequent two
photon pulse which causes a transition to an internal state
appropriate for absorption imaging. The magnetic bias field is
used to control the onset of segregation for a given density, or
alternatively the density may be increased for a fixed bias
strength to produce a similar effect.

\subsection{Equations of Motion} In order to simulate the full
behaviour of this system we wish to find suitable equations of
motion to describe the system in the Hartree-Fock regime, where a
local density approximation is valid.
 Defining the Wigner amplitudes
\begin{eqnarray}
n_{j}({\bf x},{\bf p}) = \int d^3{\bf y}\langle
\psi^\dagger_{j}({\bf x}-{\bf y}/2)\psi_{j}({\bf x}+{\bf y}/2)\rangle e^{-i{\bf p}\cdot{\bf y}/\hbar},\\
f({\bf x},{\bf p}) = \int d^3{\bf y}\langle \psi^\dagger_{1}({\bf
x}-{\bf y}/2)\psi_{2}({\bf x}+{\bf y}/2)\rangle e^{-i{\bf
p}\cdot{\bf y}/\hbar},
\end{eqnarray}
so that the densities are
\begin{eqnarray}
N_j({\bf x}) = \int {d^3{\bf p}\over(2\pi\hbar)^3}\;n_{j}({\bf
x},{\bf p})\equiv \langle \psi^\dagger_{j}({\bf x})\psi_{j}({\bf
x})
\rangle,\\
F({\bf x}) = \int {d^3{\bf p}\over(2\pi\hbar)^3}\;f({\bf x},{\bf
p})\equiv \langle \psi^\dagger_{1}({\bf x})\psi_{2}({\bf x})
\rangle,
\end{eqnarray}
and using Hartree-Fock factorisation and standard Wigner function
methods \cite{Gardiner2001}, the equations of motion are written
in terms of $n({\bf x},{\bf p}) \equiv n_1({\bf x},{\bf p}) +
n_2({\bf x},{\bf p})$ and the segregation $m({\bf x},{\bf p})
\equiv n_1({\bf x},{\bf p}) - n_2({\bf x},{\bf p})$ as
\begin{eqnarray}\label{neom}
\dot{n}({\bf x},{\bf p}) & = & \left\{-({\bf p}/
m)\cdot\nabla_{\bf x}+\nabla V_{\rm eff}({\bf x})\cdot\nabla_{\bf
p}\right\}n({\bf x},{\bf p})\nonumber\\
&&-{1\over 2}\nabla V_{\rm diff}({\bf x})\cdot\nabla_{\bf
p}m({\bf x},{\bf p})\nonumber\\
&&+u_{12}\big\{\nabla F({\bf x})\cdot\nabla_{\bf p}f({\bf x},{\bf
p})^{*}+\nabla F({\bf x})^{*}\cdot\nabla_{\bf p}f({\bf x},{\bf p})
\big\},\\
\label{meom}\dot{m}({\bf x},{\bf p}) & = & \left\{-({\bf p}/
m)\cdot\nabla_{\bf x}+\nabla V_{\rm eff}({\bf
x})\cdot\nabla_{\bf p}\right\}m({\bf x},{\bf p})\nonumber\\
&&-{1\over 2}\nabla V_{\rm diff}({\bf
x})\cdot\nabla_{\bf p}n({\bf x},{\bf p})\nonumber\\
&&-{2iu_{12}\over \hbar}\big\{f({\bf x},{\bf p})F({\bf
x})^{*}-f({\bf x},{\bf p})^{*}F({\bf x})
\big\},\\
 \label{feom}
 \dot{f}({\bf x},{\bf p}) &=& \left\{-({\bf p}/
m)\cdot\nabla_{\bf x}+\nabla V_{\rm eff}({\bf x})\cdot\nabla_{\bf
p}\right\}f({\bf x},{\bf p})\nonumber\\ &&-i{V_{\rm diff}({\bf
x})\over\hbar}f({\bf x},{\bf p})\nonumber\\
&&-i{u_{12}\over\hbar}\left(m({\bf x},{\bf p})F({\bf x})-f({\bf x},{\bf p})M(\b{x})\right)\nonumber\\
&&+ {u_{12}\over 2}\nabla F({\bf x})\cdot\nabla_{\bf p}n({\bf
x},{\bf p}),
\end{eqnarray}
where the potentials are
\begin{eqnarray}
V_{\rm eff}({\bf x}) &=& {1\over 2}\bigg(V_1({\bf x})+V_2({\bf x})+3u_{12}N({\bf x})+\delta u\;M({\bf x})\bigg),\\
V_{\rm diff}({\bf x}) &=&\hbar \Delta\omega_{\rm z}({\bf
x})+\hbar\Delta\omega_{\rm c}({\bf x}).
\end{eqnarray}
and the frequencies are the coherent frequency shift
\begin{eqnarray}\Label{omegac}
\hbar \Delta\omega_{\rm c}({\bf x}) &\equiv& 2u_{22}N_2({\bf
x})-2u_{11}N_1({\bf x})+2u_{12}(N_1({\bf x})-N_2({\bf
x}))\nonumber
\\
&\simeq&-\delta u N(\b{x}),
\end{eqnarray}
and the Zeeman shift
\begin{eqnarray}\Label{omegaz}
\hbar \Delta\omega_{\rm z}({\bf x})
&\equiv&\hbar\left(\omega_2({\bf x})-\omega_1({\bf
x})\right)=V_2({\bf x})-V_1({\bf x}).
\end{eqnarray}
For the low densities employed these equations of motion are
nearly exact for the thermal gas, apart from the relaxation caused
by collisions.

\section{Spectroscopy of the initial state}\label{sec:spec}
The result of the laser excitation
is to rotate the spin wavefunction into the $|2\rangle$ subspace,
so the one-body wavefunction is transformed to
\begin{eqnarray}\Label{bg1}
 | 1{\bf }\rangle \to  t | 1{\bf }\rangle+ r | 2{\bf }\rangle ,
\end{eqnarray}
with $ |t|^2 + |r|^2 =1$.  In field theoretic language (in the
Heisenberg picture) we describe the initial state by the field
operators $ \psi_{i1}({\bf x})$ and $ \psi_{i2}({\bf x})$, where
the subscript $i$ denotes refers to the initial state before the
pulse. These transform to
\begin{eqnarray}\Label{bg2}
\psi_{i1}({\bf x}) & \to & \psi_{1}({\bf x})=\phantom{-}t
\psi_{i1}({\bf x}) + r \psi_{i2}({\bf x})
\\ \Label{bg3}
\psi_{i2}({\bf x}) & \to & \psi_2({\bf x})=-r \psi_{i1}({\bf x}) +
t \psi_{i2}({\bf x}),
\end{eqnarray}
where the condition $|t|^2+|r|^2=1$ ensures the conservation of
total number and $r$ and $t$ are chosen real.
 We denote the population densities by $
N_j({\bf x}) = \langle \psi^\dagger_{j}({\bf x})\psi_{j}({\bf
x})\rangle$, and the coherence amplitude by $F({\bf x})=\langle
\psi^\dagger_{1}({\bf x})\psi_{2}({\bf x})\rangle$. In general the
population densities transform to
\begin{eqnarray}\fl
N_{i1}({\bf x}) \to  N_1({\bf x})&=&|t|^2N_{i1}({\bf
x})+|r|^2N_{i2}({\bf x})+t^*rF_i({\bf x})+r^*tF_i^*({\bf x}),\\\fl
N_{i2}({\bf x})  \to  N_2({\bf x})&=&|r|^2N_{i1}({\bf
x})+|t|^2N_{i2}({\bf x})-t^*rF_i^*({\bf x})-r^*tF_i({\bf x}),
\end{eqnarray}
For the \rubidium\;  experiment the initial occupation of the 2
state is zero, so that $N_{i1}({\bf x })=N({\bf x})$, $N_{i2}({\bf
x})=F_{i}({\bf x})=0$ and a two photon pulse causes the
transformation
\begin{eqnarray}\Label{spec3} N_{i1}({\bf
x })&\to&N_1({\bf x}) = \phantom{-}|t|^2N({\bf x}),
\\\Label{spec4}
N_{i2}({\bf x })&\to&N_2({\bf x})= \phantom{-}|r|^2N({\bf
x}),\\\Label{spec5} \phantom{-}F_{i}({\bf
x})&\to&\phantom{2}F({\bf x})=-t^*r N({\bf x}),\\\Label{spec6}
|F_{i}({\bf x})|^2&\to&|F({\bf x})|^2=N_1({\bf x})N_2({\bf x}),
\end{eqnarray}
so that (\ref{spec6}) corresponds to a fully coherent superposition.
\subsection{Energy Density}
 If we consider the total energy density we have in general
\begin{eqnarray}\fl\Label{Udensity}
U({\bf x}) &\equiv&\left\langle\psi_1\left(-{\hbar^2\nabla^2\over
2m}+V_1({\bf
x})\right)\psi_1\right\rangle+\left\langle\psi_2\left(-{\hbar^2\nabla^2\over
2m}+V_2({\bf x})\right)\psi_2\right\rangle\nonumber\\\fl
&&+\left\langle {u_{11}\over
2}\psi^\dagger_1\psi^\dagger_1\psi_1\psi_1
 + {u_{12}}\psi^\dagger_1\psi^\dagger_2\psi_1\psi_2
 + {u_{22}\over
 2}\psi^\dagger_2\psi^\dagger_2\psi_2\psi_2\right\rangle.
\end{eqnarray}

Changing the internal state of an atom does not change its
velocity so the kinetic energy terms may be ignored in finding the
change in energy of the gas during the transition. Using
Hartree-Fock factorization and (\ref{spec3}-\ref{spec6}), the
energy density is
\begin{eqnarray}\label{Umfeqgen}
U({\bf x})&=&u_{11}N_1({\bf x})^2+u_{12}(|F({\bf x})|^2 + N_1({\bf
x})N_2({\bf x}))+u_{22}N_2({\bf x})^2\nonumber\\
&&+V_1({\bf x})N_1({\bf x})+V_2({\bf x})N_2({\bf x}).
\end{eqnarray}
Evaluating the change in energy density for an arbitrary rotation defined by (\ref{bg2}), (\ref{bg3}) leads to
\begin{eqnarray}\label{Umfeq}
\Delta U({\bf x})&=&\bigg(u_{11}(|t|^4-1)+2u_{12}|t|^2|r|^2+u_{22}|r|^4\bigg)N({\bf x})^2\nonumber\\
&&+(V_2-V_1)|r|^2N({\bf x}).
\end{eqnarray}
\subsubsection{Infinitesimal rotation} For a very small rotation $|r|^2\ll 1 $ and
$1-|t|^4\simeq 2|r|^2$. Putting $N_2({\bf x})=\delta N({\bf x})\ll
N({\bf x})$, $|r|^4\simeq 0$ and $N_1({\bf x})\simeq N({\bf x})$,
we find
\begin{eqnarray}
\Delta U({\bf x})&=&\delta N({\bf x})\bigg(2(u_{12}-u_{11})N({\bf
x})+V_2({\bf x})-V_1({\bf x})\bigg),
\end{eqnarray}
 For \rubidium\;  this becomes (using (\ref{intapp1},\;\ref{intapp2}))
\begin{eqnarray}\fl
\Delta U({\bf x})&=&\delta N({\bf x})\bigg(-\delta u\;N({\bf
x})+V_2({\bf x})-V_1({\bf x})\bigg)=\delta N({\bf x})V_{\rm
diff}({\bf x}),
\end{eqnarray}
so that the change in energy density varies linearly with the
number of atoms transferred to the higher energy state, and is
proportional to the differential potential.
\subsubsection{$\pi/2$ pulse} For a $\pi/2$ pulse the energy density change is
\begin{eqnarray}
\Delta U({\bf x})&=&\bigg(3u_{11}+2u_{12}+u_{22}\bigg){N({\bf
x})^2\over 4}+(V_2({\bf x})-V_1({\bf x})){N({\bf x})\over2}.
\end{eqnarray}
 For \rubidium\;  this becomes
\begin{eqnarray}
\Delta U({\bf x})&=&\bigg(3u_{12}+\delta u\bigg){N({\bf x})^2\over
2}+(V_2({\bf x})-V_1({\bf x})){N({\bf x})\over2}.
\end{eqnarray}
Thus the effect of an intense pulse is to produce a nonlinear
change in the energy density via the mean field interactions.
\subsection{Coherence energy}
The analysis is simplified by noting that because the only net
force in the experiment is a weak differential force between the
two internal states, the total atomic density is approximately
conserved. The experimental data suggests that this is a
reasonable approximation \cite{Lewandowski}, and our detailed
simulations confirm this expectation. Treating $N(\bf{x})$ as
constant allows insight into the energy conserving processes in
the dynamics. In terms of the normalized moments
$\av{\b{p}_j^k(\b{x})}=\int
d^3\b{p}\;\b{p}^kn_j(\b{x},\b{p})/N_j(\b{x})(2\pi\hbar)^3$ we have
the relations
\eq{N_{1}(\b{x})\av{\b{p}_1(\b{x})}+N_{2}(\b{x})\av{\b{p}_2(\b{x})}=N(\b{x})\av{\b{p}(\b{x})}=0}
(which is essentially momentum conservation in the absence of any
external perturbation of the stationary density profile), and for
the total kinetic energy density
\begin{eqnarray}\Label{kindensity}
{N_{1}(\b{x})\av{\b{p}_1^2(\b{x})}\over
2m}+{N_{2}(\b{x})\av{\b{p}_2^2(\b{x})}\over2m}={N(\b{x})\av{\b{p}^2(\b{x})}\over2m}.
\end{eqnarray}
Here only $N(\b{x})$ is constant since the other moments and
densities will change during the relative motion of the two
species. We may now write (\ref{Umfeqgen}) in the form
\begin{eqnarray}\label{Umfdeg}
U({\bf x})&=&{3u_{12}\over4}N({\bf x})^2+{\left(V_1({\bf x})+V_2({\bf x})\right)\over 2} N({\bf
x})\\&&+{u_{12}\over4}M({\bf x})^2+{\delta u\over2} N({\bf
x})M({\bf x})-{1\over 2}\hbar\Delta\omega_{\rm z}({\bf x})M({\bf
x})\\
 &&+u_{12}|F({\bf x})|^2\\
 &&+N(\b{x}){\av{\b{p}^2(\b{x})}\over2m},
\end{eqnarray}
where the spatial segregation is
$M(\b{x})\equiv\int\;d^3\b{p}\;m(\b{x},\b{p})/(2\pi\hbar)^3$. The
first line is constant in time, and the second line only varies
with $M(\b{x})$. The third line is the coherence energy density which may
change during the motion, while the last line is a function of the
local densities and temperatures through (\ref{kindensity}). It is
clear that the coherence energy moves between $F(\b{x})$ and
$M(\b{x})$ via relative motion which changes the local temperature
and momentum of each distribution.
\section{The Ramsey frequency and the interaction energy
shift}\label{sec:Ramsey} The Ramsey technique has recently been
shown to be a particularly useful tool for exploring the coherence
properties of ultracold dilute gases
\cite{Lewandowski,Harber02,McGuirk02}. In applying the technique
to the ultra-cold non-condensed gases of these experiments there
remains an issue in the interpretation of the technique which we
wish to resolve. The problem is whether or not the Ramsey
frequency is sensitive to the decoherence of the superposition of
internal states required to resolve the Ramsey fringes. The Ramsey
technique has been developed and used extensively for probing the
relative energies of internal states in non-interacting
atomic beams. When interactions are negligible the measured
quantity is the energy difference between the two internal states
of the superposition. We will see that for the interacting case,
this is only partially true because interactions produce
a transient mean field energy which decays with the coherence of the
superposition.
\subsection{Experiment}
The experiment measures the frequency of oscillation for the
relative occupation of the two internal states, after a second
$\pi/2$ pulse is applied to the system. The final two photon pulse
transforms the fields to
\begin{eqnarray}
\psi_1\rightarrow {\psi_1+\psi_2\over\sqrt{2}},\\
\psi_2\rightarrow {\psi_2-\psi_1\over\sqrt{2}}.
\end{eqnarray}
In terms of the densities
$N_i(\b{x})\equiv\av{\psi^\dag_i\psi_i}$, and the coherence
amplitude $F(\b{x})\equiv\av{\psi_1^\dag\psi_2}$, the ratio of the
densities may be written
\begin{eqnarray}
{N_2(\b{x})\over N_1(\b{x})}={N(\b{x})-2{\rm Re\;}F(\b{x})\over
N(\b{x})+2{\rm Re\;}F(\b{x})},
\end{eqnarray}
where $N(\b{x})=N_1(\b{x}) + N_2(\b{x})$ may be assumed invariant. The ratio will then
only depend on $F(\b{x})$; in particular, it will oscillate at the
same frequency. Applying the second $\pi/2$ pulse at different
times and measuring $N_1(\b{x})/N_2(\b{x})$ thus determines the frequency of
$F(\b{x})$, and this is the frequency accessible via the Ramsey
technique of \cite{Lewandowski,McGuirk02}.
\subsection{Coherence equation} In a typical
experiment \cite{Lewandowski,McGuirk02}, a $\pi/2$ pulse generates
an equal superposition of the two hyperfine states which is then
allowed to evolve without coupling until a second $\pi/2$ pulse is
applied. We are primarily interested in the influence of the
interactions on the Ramsey frequency, and after the first pulse
the Heisenberg equations of motion generated by the interaction
terms in the Hamiltonian (\ref{Hamil}) are
\begin{eqnarray}
i\hbar \dot{\psi}_1 &=&u_{11}\psi^{\dag}_1 \psi_1\psi_1 +
u_{12}\psi^\dag_2\psi_2\psi_1,\\ i\hbar \dot{\psi}_2
&=&u_{22}\psi_2^\dag\psi_2\psi_2 + u_{12}\psi^\dag_1\psi_1\psi_2.
\end{eqnarray}
The operators $\psi_1^\dag\psi_1$ and $\psi_2^\dag\psi_2$ commute
with the Hamiltonian when $E(\b{x})= 0$, so the occupations are
preserved. The equation of motion for the coherence is
\begin{eqnarray}
i\hbar{d \av{\psi_1^\dag\psi_2}\over
dt}&=&u_{22}\av{\psi_1^\dag\psi_2^\dag\psi_2\psi_2}+u_{12}\av{\psi_1^\dag\psi_1^\dag\psi_1\psi_2}\nonumber\\
&-&u_{11}\av{\psi_1^\dag\psi_1^\dag\psi_1\psi_2}-u_{12}\av{\psi_1^\dag\psi_2^\dag\psi_2\psi_2}.
\end{eqnarray}
\subsection{Four point averages}
There are two equivalent ways of treating the four point averages
when the system is noncondensed and thermal, both of which lead to
results of the form
\begin{eqnarray}
\av{\psi_1^\dag\psi_1^\dag\psi_1\psi_2}&=& 2N_1(\b{x})F(\b{x}).
\end{eqnarray}
\begin{itemize}
\item[i)]
When the initial quantum state is nondegenerate, each field may be
written in terms of a set of orthonormal single particle
wavefunctions $\phi_r(\b{x})$
\begin{eqnarray}
\psi_i=\sum_r \phi_r(\b{x})a_{ir},
\end{eqnarray}
where $[a_{ir},a_{jk}^\dag]=\delta_{ij}\delta_{rk}$. The densities
and coherence then read
\begin{eqnarray}
\av{\psi^\dag_i\psi_j}=\sum_{r,s}\phi_r(\b{x})^*\phi_s(\b{x})\av{a^\dag_{ir}a_{js}}.
\end{eqnarray}
The eigenvalues of the operators $N_{ir}=a_{ir}^\dag a_{ir}$ which
contribute significantly are either 0 or 1. We can describe the
situation by the averages
\begin{eqnarray}
\av{a_{ir}^\dag a_{ir}}&\equiv& \bar{N}_{ir},\\
\av{a_{1r}^\dag a_{2r}}&\equiv& \bar{M}_r.
\end{eqnarray}
For all averages $\av{a_{ir}^\dag a_{js}}=0$ when $r\neq s$,
corresponding to independently occupied modes. We then have
\begin{eqnarray}
\av{\psi_1^\dag\psi_2}=\sum_{r}|\phi_r(\b{x})|^2\bar{M}_r,
\end{eqnarray}
and the four point averages become, for example
\begin{eqnarray}\fl
\av{a_{1r}^\dag a_{1s}^\dag a_{1j}
a_{2k}}&=&\cases{&\hskip -6mm
$\delta_{rj}\delta_{sk}\bar{N}_{1r}\bar{M}_s
+\delta_{sj}\delta_{rk}\bar{N}_{1s}\bar{M}_r$\;\;when $r\neq s$ and $j\neq k$,
\\&
\hskip -6mm
 0\;\;otherwise.\\}
\end{eqnarray}
The four point average becomes
\begin{eqnarray}\label{fieldav}
\av{\psi_1^\dag\psi_1^\dag\psi_1\psi_2}&=&2\sum_{r,s, \;r\neq
s}|\phi_r(\b{x})|^2|\phi_s(\b{x})|^2\bar{N}_{1r}\bar{M}_s,\nonumber\\
&\simeq& 2N_1(\b{x})F(\b{x}),
\end{eqnarray}
where the approximation is valid if the occupation is spread over
very many modes, as for a thermal state, so that the term with
$r=s$ is negligible.
\item[ii)]
Alternatively, the same result may be found using Hartree-Fock
factorization for the averages over the field operators. This
method uses the Gaussian statistics of the thermal gas, which may
hold under more general circumstances than the arguments of i),
but is equivalent for the system under consideration. The averages
are
\begin{eqnarray}\fl
\av{\psi_1^\dag\psi_2^\dag\psi_3\psi_4}=\av{\psi_1^\dag\psi_3}\av{\psi_2^\dag\psi_4}
+\av{\psi_1^\dag\psi_4}\av{\psi_2^\dag\psi_3}+\av{\psi_1^\dag\psi_2^\dag}\av{\psi_3\psi_4}.
\end{eqnarray}
For a thermal gas there is no anomalous average, so we recover the
same result as (\ref{fieldav})
\begin{eqnarray}
\av{\psi_1^\dag\psi_1^\dag\psi_1\psi_2}=2N_1(\b{x})F(\b{x}).
\end{eqnarray}
\end{itemize}
\subsection{Ramsey frequency}
Using either approach, the equation of motion for the coherence
becomes
\begin{eqnarray}\fl\label{rams}
i\hbar{d\over
dt}F(\b{x})&=&\left[2u_{22}N_2(\b{x})-2u_{11}N_1(\b{x})+2u_{12}(N_1(\b{x})-N_2(\b{x}))\right]F(\b{x})\nonumber\\\fl
&\equiv&\hbar\omega_R(\b{x})F(\b{x}).
\end{eqnarray}
Thus if the gas is thermal, the frequency of oscillation
$\omega_R(\b{x})$ is independent of the amplitude $|F(\b{x})|$. In
particular if some kind of damping reduces $|F(\b{x})|$ with time,
this does {\it not} change the Ramsey frequency.
\subsubsection{Transport and trap effects}
The full equation of motion (\ref{feom}) may be integrated to find
\begin{eqnarray}
{\partial F({\bf x})\over \partial t}+\nabla\cdot(F({\bf x}){\bf
v}_{F}({\bf x}))&=&-i\left(\Delta\omega_{c}({\bf
x})+\Delta\omega_{\rm z}({\bf x})\right)F({\bf
x})\nonumber\\\Label{Fdot} &\equiv&-i\omega_{R}(\b{x})F({\bf x})
\end{eqnarray}
where the velocity is
\begin{eqnarray}
{\bf v}_{F}(\b{x})={1\over m  F({\bf x})}\int {d^3{\bf
p}\over(2\pi\hbar)^3}{\bf p}f({\bf x},{\bf p}).
\end{eqnarray}
When the velocity vanishes we recover (\ref{rams}) with an
additional shift caused by the trap energy difference for the two
states. The coherence current described by $\b{v}_F$ may alter the
measured Ramsey fringes, and indeed full simulations may be
required for comparison with experiment when significant motion occurs.
\subsection{Interaction energy shift} The quantity which is
sensitive to the loss of coherence is the interaction energy,
which in general does not correspond to the Ramsey frequency
$\omega_R(\b{x})$.

The interaction energy change caused by changing the internal
state from $\ket{1}$ to $\ket{2}$ is found from the chemical
potentials associated with this change
\begin{eqnarray}
\mu_j({\bf x})&\equiv&{\partial U({\bf x})\over\partial N_j({\bf
x})}.
\end{eqnarray}
Neglecting the transport and trap terms for simplicity, we use the mean energy density for the interactions
\begin{eqnarray}
U(\b{x})={u_{11}\over
2}\av{\psi_1^\dag\psi_1^\dag\psi_1\psi_1}+u_{12}\av{\psi_1^\dag\psi_2^\dag\psi_1\psi_2}+{u_{22}\over
2}\av{\psi_2^\dag\psi_2^\dag\psi_2\psi_2},
\end{eqnarray}
and factorize averages as in (\ref{fieldav}), to find
\begin{eqnarray}\label{mu1}
\mu_1&=&2u_{11}N_1 + u_{12}N_2+{\partial |F|^2\over\partial
N_1},\\\label{mu2}
 \mu_2&=&2u_{22}N_2+u_{12}N_1+{\partial
|F|^2\over\partial N_2}.
\end{eqnarray}
The change in energy caused by this transition is $\Delta\mu
=\mu_2-\mu_1\equiv\hbar\Delta\omega_{\mu}$, so that using
(\ref{mu1}), (\ref{mu2}) we find
\begin{eqnarray}\label{deltamu}
\hbar\Delta\omega_{\mu} &=&2u_{22}N_2-2u_{11}N_1+u_{12}(N_1-N_2)\nonumber\\
&&+u_{12}{\partial |F|^2\over\partial N_2}-u_{12}{\partial
|F|^2\over\partial N_1}.
\end{eqnarray}
The derivatives of the coherence energy density are not evaluated
because it is not usually possible to do so in any direct manner
since $|F(\b{x})|^2$ cannot generally be specified by knowledge of
the $N_i(\b{x})$. The Cauchy-Schwartz inequality leads to
\begin{eqnarray}
|F(\b{x})|^2\leq N_1(\b{x})N_2(\b{x}).
\end{eqnarray}
When $|F(\b{x})|^2=N_1(\b{x})N_2(\b{x})$ the superposition is
purely coherent, leading to a factor of 2 in the cross interaction
part of the frequency shift; whereas for $|F|^2=0$ the extra
factor due to the coherence is absent and we recover the thermal
result. A simple model which we will use is found by taking
$|F|^2=\alpha(t) N_1N_2$, with $\alpha(0)=1$ reconstructing the
initial condition (\ref{spec6}), and $\alpha(t\to\infty)=0$ medelling the
damping to thermal equilibrium. The mean field energy shift becomes
\begin{eqnarray}
\hbar\Delta\omega_\mu=2u_{22}N_2-2u_{11}N_1+2\alpha(t)u_{12}(N_1-N_2)
\end{eqnarray}
 Immediately after the pulse
\begin{eqnarray}\label{deltamudef}
\Delta\omega_{\mu}(\b{x}) = \omega_{R}(\b{x}),
\end{eqnarray}
so that for short times the Ramsey frequency coincides with the
interaction energy shift. Inclusion of the effect of collisions
will cause $|F(\b{x})|$ to decay with time. The simplest way of
doing this is to consider the case where there is no differential
potential gradient so that segregation cannot occur. The
collisions will still have a relaxation effect on the coherence
and as a first approximation we may use the equation of motion
\begin{eqnarray}
\dot{F}=-i\omega_R(\b{x})F-{1\over\tau_{c}}F,
\end{eqnarray}
where $\tau_c$ is the timescale of relaxation. Solving this and
using (\ref{deltamu}) gives
\begin{eqnarray}
\hbar\Delta\omega_{\mu}= 2u_{22}N_2-2u_{11}N_1+u_{12}(N_1-N_2)(1+\e^{-2t/\tau_c}).\nonumber\\
\end{eqnarray}
Clearly the mean field energy shift {\it is} sensitive to the
decay of coherence, and decays at twice the rate for the
amplitude.

\section{Kinetic Theory of Coherence Damping}\Label{sec:cdamp}
We now consider the role of collisions in dissipating coherence.
Our starting point is the quantum kinetic theory of Gardiner and
Zoller \cite{QKI}, which may be used to find an expression for the
damping rate of $f({\bf x},{\bf p})$. While the expressions found
are quite explicit, they are not easy to simulate, so that for the
actual simulations we will use simplified forms based on a
relaxation time approximation.

\subsection{Quantum Kinetic Theory with internal degrees of
freedom} In order to derive the damping for $f(\b{x},\b{p})$ we
first consider the damping term in the equation of motion for the
 reduced density matrix for a gas with a single internal
 state. For brevity of notation we suppress $\b{x}$ arguments where possible
 since the damping is local. For ease of comparison with \cite{QKI} we carry out the
 following in $\b{K}$ variables.
We seek the equation of motion for the terms
\begin{eqnarray}\Label{nlmdef}
f_{lm}(\b{x},\b{k})=\int
d^3\b{y}\av{\psi_l^{\dagger}(\b{x}-\b{y}/2)\psi_m(\b{x}+\b{y}/2)}\e^{-i\b{k}\cdot\b{y}}
\end{eqnarray}
where $l,m=1,2$. The derivation of the Quantum Boltzmann equation
may then be carried out along similar lines to that in \cite{QKI}.
The principal modification of the derivation is due to the
different scattering allowed between distinguishable internal
states of the particles, and the preservation of possible
coherence between internal states. The resolution of the field
operator for internal state $i$ into its momentum components
$\psi_{i\b{K}}(\b{x})$ is defined similarly to (31) of \cite{QKI}
as
\begin{eqnarray}
\psi_i(\b{x})=\sum_{\b{K}}\e^{-i\b{K}\cdot\b{x}}\psi_{i\b{K}}(\b{x}).
\end{eqnarray}
The projector defined in (48) of \cite{QKI} for the derivation of
the master equation simplifies slightly because no Bose-Einstein
condensate is present in this case, and becomes
\begin{eqnarray}
p_{\b{N}}\ket{\b{n}}=\cases{&$\ket{\b{n}}$\;\;if\;\;
$\sum_{\b{r},i}n_i(\b{K},\b{r})=N(\b{K})$\\
&$0$\;\; otherwise.\\}
\end{eqnarray}
This projector identifies all configurations with the same
distribution in $\b{K}$, but leaves the position and internal
state distributions undisturbed. To evaluate the effect of
collisions on the coherence we require the equation of motion for
the reduced density matrix
\begin{eqnarray}
{\cal P}_{\b{N}}\rho=p_{\b{N}}\rho p_{\b{N}}\equiv v_{\b{N}},
\end{eqnarray}
induced by the interaction Hamiltonian
\begin{eqnarray}
H_I={1\over2}\sum_{ij,\;\b{e}}U_{ij}(1,2,3,4),
\end{eqnarray}
in which we have used the notation for the sum over all momenta
defined in \cite{QKI}
\begin{eqnarray}
\sum_\b{e}\equiv\sum_{\b{K}_1,\b{K}_2,\b{K}_3,\b{K}_4},
\end{eqnarray}
and the interaction operator $U_{ij}(1234)=U_{ij}(\b{e})$:
\begin{eqnarray}
U_{ij}(1234)&=&\int d^3\b{x}\int
d^3\b{x^{\prime}}\;\e^{(i\b{K}_1\cdot\b{x}+i\b{K}_2\cdot\b{x^{\prime}}-i\b{K}_3\cdot\b{x}-i\b{K}_4\cdot\b{x^{\prime}})}\nonumber\\
&&\times\psi^{\dagger}_{i\b{K}_1}(\b{x})\psi^{\dagger}_{j\b{K}_2}(\b{x^{\prime}})u_{ij}(\b{x}-\b{x^{\prime}})\psi_{i\b{K}_3}(\b{x^{\prime}})\psi_{j\b{K}_4}(\b{x}).
\end{eqnarray}
In this form $H_I$ is just the $\b{K}$ space representation of the
second line of (\ref{Hamil}).
 In what follows we will make the usual psuedopotential
approximation for the interaction parameters and set
\begin{eqnarray}
u_{ij}(\b{x}-\b{x^{\prime}})={4\pi\hbar^2a_{ij}\over
m}\delta(\b{x}-\b{x^{\prime}})\equiv
u_{ij}\delta(\b{x}-\b{x^{\prime}}).
\end{eqnarray}
In order to describe the exchange collision $1,2\to 4,3$ we also
define the interaction operator $U_{ij}(1243)=U_{ij}(\b{\bar{e}})$
as
\begin{eqnarray}
U_{ij}(\b{\bar{e}})&=&u_{ij}\int d^3\b{x}
\;\e^{i(\b{K}_1+\b{K}_2-\b{K}_3-\b{K}_4)\cdot\b{x}}\nonumber\\
&&\times\psi^{\dagger}_{i\b{K}_1}(\b{x})\psi^{\dagger}_{j\b{K}_2}(\b{x})\psi_{j\b{K}_3}(\b{x})\psi_{i\b{K}_4}(\b{x})
\end{eqnarray}
which reverses the momenta of the last two operators.
 Carrying out the
procedures of \cite{QKI} leads to the density matrix collision
term similar to (75e) of \cite{QKI}
\begin{eqnarray}\fl\Label{veom}
\dot{v}_{\b{N}}(t)|_{\rm coll}&=&{\pi\over2\hbar^2}\sum_{ijIJ,\;
\b{e}}\;\delta(\Delta\omega(\b{e}))\nonumber\\\fl
&&\times\Bigg\{\;\;\left[2U_{ij}(\b{e})v_{\b{N-e}}
(t)U_{IJ}^{\dagger}(\b{e})-U_{ij}(\b{e})U_{IJ}^{\dagger}(\b{e})v_{\b{N}}(t)-v_{\b{N}}(t)U_{IJ}^{\dagger}(\b{e})U_{ij}(\b{e})\right]\nonumber\\\fl
&&\;\;\;\;\;\;+\Big[U_{ij}(\b{e})v_{\b{N-e}}
(t)U_{IJ}^{\dagger}(\b{\bar{e}})+U_{IJ}(\b{\bar{e}})v_{\b{N-e}}(t)U_{ij}^{\dagger}(\b{e})\nonumber\\\fl
&&\;\;\;\;\;\;\;\;\;\;\;\;\;-U_{ij}(\b{e})U_{IJ}^{\dagger}(\b{\bar{e}})v_{\b{N}}(t)-v_{\b{N}}(t)U_{IJ}^{\dagger}(\b{\bar{e}})U_{ij}(\b{e})\Big]\;\Bigg\},
\end{eqnarray}
where $\Delta\omega(\b{e})=\omega_4+\omega_3-\omega_2-\omega_1$.
 The first line is the contribution from the collisions
$1,2\to3,4$ and $2,1\to4,3$ and is a sum of terms like (75e) for a
single internal state , while the remaining terms describe the
collisions $1,2\to4,3$ and $2,1\to3,4$. Of course if only one
internal state occurs, the substitution $i=j=I=J$ restores the
exchange symmetry of the operator
$U_{ii}(\b{e})=U_{ii}(\b{\bar{e}})=U_{ii}^{\dagger}(-\b{e})$ and
we recover the ordinary collision term.
\subsection{Modified Quantum Boltzmann equation}
We now seek the Quantum Boltzmann equation for this system, which
is the equation of motion for (\ref{nlmdef}) in the continuum
limit. Tracing over the reduced density matrix allows us to write
the equation of motion for
\begin{eqnarray}
\av{\psi^{\dagger}_{l\b{k}}\psi_{m\b{k}}}= \sum_{\b{N}}{\rm
Tr}\{v_{\b{N}}\psi^{\dagger}_{l\b{k}}\psi_{m\b{k}}\}
\end{eqnarray}
in the form
\begin{eqnarray}\fl
{d\over dt}
\av{\psi^{\dagger}_{l\b{k}}\psi_{m\b{k}}}&=&\left({\pi\over2\hbar^2}\right)\sum_{ijIJ,\;
\b{e}}\;\delta(\Delta\omega(\b{e}))\nonumber\\\fl\Label{directav}
&&\times\Bigg(\;\;\;\av{U_{IJ}^{\dagger}(\b{e})[\psi^{\dagger}_{l\b{k}}\psi_{m\b{k}},U_{ij}(\b{e})]}
+\av{[U_{ij}(\b{e}),\psi^{\dagger}_{l\b{k}}\psi_{m\b{k}}]U_{IJ}^{\dagger}(\b{e})}
\\\fl\Label{exchangeav}
&&\;\;\;\;\;\;+\av{U_{IJ}^{\dagger}(\b{\bar{e}})[\psi^{\dagger}_{l\b{k}}\psi_{m\b{k}},U_{ij}(\b{e})]}
+\av{[U_{ij}(\b{e}),\psi^{\dagger}_{l\b{k}}\psi_{m\b{k}}]U_{IJ}^{\dagger}(\b{\bar{e}})}\;\;\;\Bigg).
\end{eqnarray}
The terms in lines (\ref{directav}) and (\ref{exchangeav}) give
the direct and exchange scattering contributions respectively. The
details of passing to the continuum limit are discussed in
\ref{A1}.  For the purposes of ease of comparison with other
calculations of this kind and coherence with the rest of this
paper, we will also write the final result in terms of momenta
$\b{P}=\hbar\b{K}$ and energies $\epsilon=\hbar\omega$ instead of
frequencies and wave vectors. We eventually find
\begin{eqnarray}\fl\Label{collisionintegral}
&&{\partial f_{lm}\over\partial t }\Big|_{\rm coll
}={\pi\over\hbar}\int\;{d^3\b{P}_2\over(2\pi\hbar)^3}\int\;{d^3\b{P}_3\over(2\pi\hbar)^3}\int\;d^3\b{P}_4
\nonumber\\\fl
&&\;\;\;\;\;\;\;\;\;\;\;\;\;\;\times\delta(\b{P}+\b{P}_2-\b{P}_3-\b{P}_4)\delta(\epsilon+\epsilon_2-\epsilon_3-\epsilon_4)\sum_{iIJ}u_{IJ}\nonumber\\\fl
&&\times\Bigg[\Bigg((f_{lJ}(\b{P})+\delta_{lJ})(f_{iI}(\b{P}_2)+\delta_{iI})f_{Im}(\b{P}_3)f_{Ji}(\b{P}_4)\nonumber\\\fl
&&\;\;\;\;\;\;\;\;\;\;\;\;\;\;\;\;\;\;-f_{lJ}(\b{P})f_{iI}(\b{P}_2)(f_{Im}(\b{P}_3)+\delta_{Im})(f_{Ji}(\b{P}_4)+\delta_{Ji})\nonumber\\\fl\nonumber\\\fl
&&\;\;\;\;\;\;\;\;\;+(f_{lJ}(\b{P})+\delta_{lJ})(f_{iI}(\b{P}_2)+\delta_{iI})f_{Ii}(\b{P}_3)f_{Jm}(\b{P}_4)\nonumber\\\fl
&&\;\;\;\;\;\;\;\;\;\;\;\;\;\;\;\;\;\;-f_{lJ}(\b{P})f_{iI}(\b{P}_2)(f_{Jm}(\b{P}_3)+\delta_{Jm})(f_{Ii}(\b{P}_4)+\delta_{Ii})\Bigg)u_{im}\nonumber\\\fl
&&\;\;\;\;+\Bigg((f_{Jm}(\b{P})+\delta_{Jm})(f_{Ii}(\b{P}_2)+\delta_{Ii})f_{lI}(\b{P}_3)f_{iJ}(\b{P}_4)\nonumber\\\fl
&&\;\;\;\;\;\;\;\;\;\;\;\;\;\;\;\;\;\;-f_{Jm}(\b{P})f_{Ii}(\b{P}_2)(f_{lI}(\b{P}_3)+\delta_{lI})(f_{iJ}(\b{P}_4)+\delta_{iJ})\nonumber\\\fl\nonumber\\\fl
&&\;\;\;\;\;\;\;\;\;+(f_{Jm}(\b{P})+\delta_{Jm})(f_{Ii}(\b{P}_2)+\delta_{Ii})f_{iI}(\b{P}_3)f_{lJ}(\b{P}_4)\nonumber\\\fl
&&\;\;\;\;\;\;\;\;\;\;\;\;\;\;\;\;\;\;-f_{Jm}(\b{P})f_{Ii}(\b{P}_2)(f_{lJ}(\b{P}_3)+\delta_{lJ})(f_{iI}(\b{P}_4)+\delta_{iI})\Bigg)u_{il}\Bigg]\\\nonumber\fl
\end{eqnarray}
This is a sum of direct and exchange Quantum Boltzmann collision
terms, summed over the possible interactions for each case.
 This is now in a form where simplifications may be introduced,
and there are a number of useful results that can be obtained from
this equation for the particular case of \rubidium.
\subsection{The case of equal scattering lengths and low density}
 If we drop the terms proportional to
$\delta_{ij}$, and set the scattering lengths equal, so that
$u_{ij}=u_{12}$, we straightforwardly recover the Boltzmann limit
for $l\neq m$, in agreement with (6) of \cite{Williams2002},
\begin{eqnarray}\fl\Label{Boltzlim}
{\partial f_{lm}\over\partial t }\Big|_{\rm 2 }&=&{\pi
u_{12}^2\over\hbar}\int\;{d^3\b{P}_2\over(2\pi\hbar)^3}\int\;{d^3\b{P}_3\over(2\pi\hbar)^3}\int\;d^3\b{P}_4
\nonumber\\\fl
&&\times\delta(\b{P}+\b{P}_2-\b{P}_3-\b{P}_4)\delta(\epsilon+\epsilon_2-\epsilon_3-\epsilon_4)\nonumber\\\fl
&&\times\left\{3n(\b{P}_3)f_{lm}(\b{P}_4)+n(\b{P}_4)f_{lm}(\b{P}_3)-n(\b{P})f_{lm}(\b{P}_2)-3n(\b{P}_2)f_{lm}(\b{P})\right\},\nonumber\\\fl
\end{eqnarray}
where we have set $f_{11}+f_{22}=n$.  These terms involve products
of two distributions, and we will use a notation in which terms
involving products of $j$ distributions will be denoted by
$\dot{f}_{lm}|_{j}$.

It is important to note that the presence of coherences in the
collision integrals means that the usual arguments for the
relaxation time approximation are not strictly applicable. For the
$\ket{1}$ state for example, with the same approximations as
above, the collision term becomes
\begin{eqnarray}\fl
{\partial f_{11}\over\partial t }\Big|_{\rm 2 }&=&{\pi
u_{12}^2\over\hbar}\int\;{d^3\b{P}_2\over(2\pi\hbar)^3}\int\;{d^3\b{P}_3\over(2\pi\hbar)^3}\int\;d^3\b{P}_4
\nonumber\\\fl
&&\times\delta(\b{P}+\b{P}_2-\b{P}_3-\b{P}_4)\delta(\epsilon+\epsilon_2-\epsilon_3-\epsilon_4)\nonumber\\\fl
&&\times\Big\{2n(\b{P}_3)f_{11}(\b{P}_4)-2n(\b{P}_2)f_{11}(\b{P})\nonumber\\\fl
&&\;\;\;+2f_{11}(\b{P}_3)f_{11}(\b{P}_4)-2f_{11}(\b{P}_2)f_{11}(\b{P})\nonumber\\\fl
&&\;\;\;+2f_{12}(\b{P}_3)f_{21}(\b{P}_4)-f_{12}(\b{P})f_{21}(\b{P}_2)-f_{12}(\b{P}_2)f_{21}(\b{P})\Big\},
\end{eqnarray}
so that we expect the coherence terms, like those in the last
line, to provide extra damping for the distributions.
\subsection{Equilibrium coherence damping}
The Quantum Boltzmann collision integrals for the distributions
$n_i(\b{x},\b{p})$ vanish if the distributions are in local
equilibrium. When there are two interacting internal states with
different S-wave scattering lengths, there is an additional
damping effect caused by the cross interactions. In the experiment
the degeneracy is typically about $10\%$ so we will find the effect of
keeping the differences in scattering lengths in
(\ref{collisionintegral}) using the Boltzmann equilibrium momentum
distribution. Since the quartic terms in (\ref{collisionintegral})
cancel, there are two processes to consider. The quadratic terms
provide the most important contribution for low phase space
density. The cubic rate has the additional interesting feature of
depending on the phase space density $N(\b{x})\lambda_{\rm th }^3$,
where the thermal deBroglie wavelength is $\lambda_{\rm th}\equiv
(2\pi\hbar^2/mk_BT)^{1/2}$.
\subsubsection{Quadratic terms} Retaining the differences in
scattering lengths we find the term corresponding to the Boltzmann
approximation (\ref{Boltzlim}) is
\begin{eqnarray}\fl
{\partial f_{lm}\over\partial t }\Big|_{\rm 2
}&=&{\pi\over\hbar}\int\;{d^3\b{P}_2\over(2\pi\hbar)^3}\int\;{d^3\b{P}_3\over(2\pi\hbar)^3}\int\;d^3\b{P}_4
\nonumber\\\fl
&&\times\delta(\b{P}+\b{P}_2-\b{P}_3-\b{P}_4)\delta(\epsilon+\epsilon_2-\epsilon_3-\epsilon_4)\nonumber\\\fl
&&\times\sum_{i}\Bigg\{2u_{il}u_{im}(f_{im}(\b{P}_3)f_{li}(\b{P}_4)+f_{ii}(\b{P}_3)f_{lm}(\b{P}_4))\nonumber\\\fl
&&\;\;\;\;\;\;\;\;\;\;\;\;\;\;\;\;\;-u_{il}^2(f_{im}(\b{P})f_{li}(\b{P}_2)+f_{lm}(\b{P})f_{ii}(\b{P}_2))\nonumber\\\fl
&&\;\;\;\;\;\;\;\;\;\;\;\;\;\;\;\;\;-u_{im}^2(f_{li}(\b{P})f_{im}(\b{P}_2)+f_{lm}(\b{P})f_{ii}(\b{P}_2))\;\Bigg\}
\end{eqnarray}
To find the local damping rate of the coherence amplitude
$N_{lm}(\b{x})=\int{d^3\b{p}\over(2\pi\hbar)^3}f_{lm}(\b{x},\b{p})$ we
approximate the equilibrium distribution by the local Boltzmann
equilibrium form
\begin{eqnarray}
f_{lm}(\b{x},\b{p})=N_{lm}(\b{x})(2\pi\hbar)^3\left({\alpha\over\pi
}\right)^{3/2}\exp{\left(-\alpha \b{p}^2\right)},
\end{eqnarray}
with $\alpha=1/2mk_BT$. Integrating leads to
\begin{eqnarray}\fl
{\partial N_{lm}\over \partial t}\Big|_{2}&=&-I_{34}{\pi
\over\hbar}\sum_{i}\left(u_{il}-u_{im}\right)^2(N_{im}(\b{x})N_{li}(\b{x})+N_{ii}(\b{x})N_{lm}(\b{x})),
\end{eqnarray}
where $I_{34}$ is calculated in \ref{A1}. Clearly the collisions
between the same hyperfine state do not cause damping of coherence, since
this expression vanishes for $l=m$. When $l\neq m$, the
interaction terms are $(u_{il}-u_{im})^2=(\delta u/2)^2$ for all
$i$. In terms of the thermal relative  velocity, the final result
may be written as
\begin{eqnarray}\Label{Boltzdamp}
{\partial F(\b{x})\over \partial t}\Big|_2&=&-4\pi (\delta
a)^2\bar{v}_rN(\b{x})F(\b{x}),
\end{eqnarray}
where $\delta a = a_{11}-a_{12}$. The rate coefficient takes
the form $\sigma v \rho(\b{x})$, which occurs in S-wave scattering
of distinguishable particles, but with an effective scattering
length $\delta a$. For the experiments of \cite{Lewandowski} this produces a $1/e$ decay time for the coherence of
$\sim 13\;{\rm s}$ at the center of the trap, as noted in \cite{Williams2002}.
\subsubsection{Cubic damping}
For single species collisions, the cubic terms in the Quantum
Boltzmann equation are responsible for the Bose enhancement of
scattering which is crucial in describing BEC growth, four wave
mixing and other statistical collision phenomena that can occur in
nonlinear atom optics at high phase space density. For two internal states, the cubic terms of
(\ref{collisionintegral}) are
\begin{eqnarray}\fl\Label{cubicf}
\fl &&{\partial f_{lm}\over\partial t }\Big|_{\rm 3
}={\pi\over\hbar}\int\;{d^3\b{P}_2\over(2\pi\hbar)^3}\int\;{d^3\b{P}_3\over(2\pi\hbar)^3}\int\;d^3\b{P}_4
\nonumber\\\fl
&&\;\;\;\;\;\;\;\;\;\;\;\;\;\;\times\delta(\b{P}+\b{P}_2-\b{P}_3-\b{P}_4)\delta(\epsilon+\epsilon_2-\epsilon_3-\epsilon_4)\sum_{i\;J}\nonumber\\\fl
&&\times\Bigg[\Bigg(f_{lJ}(\b{P})f_{im}(\b{P}_3)f_{Ji}(\b{P}_4)u_{iJ}+f_{iJ}(\b{P}_2)f_{Jm}(\b{P}_3)f_{li}(\b{P}_4)u_{lJ}\nonumber\\\fl
&&\;\;\;\;\;\;\;\;\;\;\;\;\;\;\;\;\;\;\;\;\;\;\;\;-f_{Jm}(\b{P}_3)f_{li}(\b{P})f_{iJ}(\b{P}_2)u_{iJ}-f_{Ji}(\b{P}_4)f_{lJ}(\b{P})f_{im}(\b{P}_2)u_{mJ}\nonumber\\\fl\nonumber\\\fl
&&\;\;\;\;\;\;\;\;\;+f_{lJ}(\b{P})f_{ii}(\b{P}_3)f_{Jm}(\b{P}_4)u_{iJ}+f_{iJ}(\b{P}_2)f_{Ji}(\b{P}_3)f_{lm}(\b{P}_4)u_{lJ}\nonumber\\\fl
&&\;\;\;\;\;\;\;\;\;\;\;\;\;\;\;\;\;\;\;\;\;\;\;\;-f_{Jm}(\b{P}_3)f_{lJ}(\b{P})f_{ii}(\b{P}_2)u_{iJ}-f_{Ji}(\b{P}_4)f_{lm}(\b{P})f_{iJ}(\b{P}_2)u_{mJ}\Bigg)u_{im}\nonumber\\\fl
&&+\Bigg(f_{Jm}(\b{P})f_{li}(\b{P}_3)f_{iJ}(\b{P}_4)u_{iJ}+f_{Ji}(\b{P}_2)f_{lJ}(\b{P}_3)f_{im}(\b{P}_4)u_{mJ}\nonumber\\\fl
&&\;\;\;\;\;\;\;\;\;\;\;\;\;\;\;\;\;\;\;\;\;\;\;\;-f_{lJ}(\b{P}_3)f_{im}(\b{P})f_{Ji}(\b{P}_2)u_{iJ}-f_{iJ}(\b{P}_4)f_{Jm}(\b{P})f_{li}(\b{P}_2)u_{lJ}\nonumber\\\fl\nonumber\\\fl
&&\;\;\;\;\;\;\;\;\;+f_{Jm}(\b{P})f_{ii}(\b{P}_3)f_{lJ}(\b{P}_4)u_{iJ}+f_{Ji}(\b{P}_2)f_{iJ}(\b{P}_3)f_{lm}(\b{P}_4)u_{mJ}\nonumber\\\fl
&&\;\;\;\;\;\;\;\;\;\;\;\;\;\;\;\;\;\;\;\;\;\;\;\;-f_{lJ}(\b{P}_3)f_{Jm}(\b{P})f_{ii}(\b{P}_2)u_{iJ}-f_{iJ}(\b{P}_4)f_{lm}(\b{P})f_{Ji}(\b{P}_2)u_{lJ}\Bigg)u_{il}\Bigg].
\end{eqnarray}
 Using a thermal Boltzmann distribution and the same procedures as in the previous section, this may be reduced to
\begin{eqnarray}\fl
{\partial N_{lm}\over\partial t }\Big|_{\rm 3
}&=&-{\pi\over\hbar}I_{234}\sum_{i\;J}\Bigg\{u_{iJ}(u_{il}-u_{im})(N_{lJ}N_{Ji}N_{im}-N_{li}N_{iJ}N_{Jm})\nonumber\\\fl
&&+(u_{il}u_{Jl}+u_{im}u_{Jm}-2u_{im}u_{Jl})(N_{li}N_{iJ}N_{Jm}+N_{iJ}N_{Ji}N_{lm})\Bigg\}.
\end{eqnarray}
As for the quadratic rate, when $l=m$ the cubic contribution
vanishes. When $l\neq m$ we may use $I_{234}$ from \ref{A1} to
find
\begin{eqnarray}\Label{cubic}\fl
{\partial F(\b{x})\over\partial t }\Big|_{\rm 3
}&=&-\sqrt{3\over8}4\pi(\delta a)^2\bar{v}_r \lambda_{\rm th}^3
\left(N_{1}(\b{x})^2+N_{2}(\b{x})^2+2|F(\b{x})|^2\right)F(\b{x}),
\end{eqnarray}
where again we have separated the effective cross-section. For the
rubidium experiment where degeneracy is about $10\%$ we estimate
the peak damping rate by simply using $N_{ij}=N/2$ which holds
just after the first $\pi/2$ pulse. Equation (\ref{cubic}) now
becomes
\begin{eqnarray}\Label{cubicestimate}
{\partial F(\b{x})\over\partial t }\Big|_{\rm 3
}&=&-\sqrt{3\over8}4\pi(\delta a)^2\bar{v}_r \lambda_{\rm
th}^3N(\b{x}) N(\b{x})F(\b{x}).
\end{eqnarray}
This rate is proportional to $\sigma v \rho(\b{x})\lambda_{\rm
th}^3\rho(\b{x})$, which is distinguished from (\ref{Boltzdamp})
by an extra factor of the phase space density. For the JILA
experiment \cite{Lewandowski} this process leads to a $1/e$ decay
time at the trap center of $\sim 240$ s. The cubic rate is clearly
unimportant in this regime, but if the phase space density is
increased significantly, the cubic effects can approach or even
exceed the quadratic damping.

It is also interesting to note the presence of a self interaction
term in the cubic damping rate, proportional to $|F(\b{x})|^2$. It
was found in \cite{Harber02} that the coherent velocity changing
collisions responsible for spin waves may suppress the decoherence
caused by atomic motion in a non-uniform system. For high phase
space density there is also the possibility of the opposite effect
becoming important, so that even in equilibrium the coherent
interactions may cause significant additional damping.

\subsection{Damping to local equilibrium}\Label{dampingsection}
The equations of motion must be modified to include the effects of
collisions on the distributions. This causes a relaxation of the
momentum distributions towards equilibrium which will be dealt
with using a relaxation time approximation. The rates derived from
simple collision time considerations require modification to
account for the effect of coherence on the relaxation. The
coherence distribution between the two fields has a significant
effect on the relaxation process because it plays a similar role to
the single species distributions.
\subsubsection{Relaxation time approximation} The collision damping
term in the equations of motion for a single internal state is
written in the relaxation time approximation as
\begin{eqnarray}
\dot{n}({\bf x},{\bf p})|_{coll} &=&-\tau_d^{-1}(\b{x})(n({\bf
x},{\bf p})-n^{(eq)}({\bf x},{\bf p}))
\end{eqnarray}
where $n^{(eq)}({\bf x},{\bf p})$ is the local equilibrium, and
$\tau_d$ is damping time which may depend on position and the
relative velocities of the two internal states.
 An estimate for the collision rate may be found from
elementary kinetic theory if the correct quantum mechanical
scattering cross sections are used. The hard sphere model of
interactions leads to the collision frequency for a given particle
of internal state $i$ with particles of internal state $j$ and
density $N_j({\bf x})$ given by \cite{Reichl}
\begin{eqnarray}
\nu_{ij}=N_j({\bf x})\pi d_{ij}^2\langle v_r(\b{x})\rangle_{ij}
\end{eqnarray}
where $d_{ij}=(d_i+d_j)/2$ is the average hard sphere particle
diameter and $\langle v_r(\b{x})\rangle_{ij}$ is the mean relative
velocity of particles $i$ and $j$. If the masses are equal, that
is $m_i=m_j\equiv m$, then in the absence of any net relative
velocity for each state, the velocity assumes the thermal value
\begin{eqnarray}\label{vtherm}
\bar{v}_r&\equiv&\sqrt{16k_BT/\pi m}.
\end{eqnarray}
 The quantum formulation of
the cross section is then given by the substitution $\pi
d_{ij}^2=\sigma_{ij}$, where $\sigma_{ij}=4\pi
a_{ij}^2(1+\delta_{ij})$. It is convenient to define the local
equilibrium distributions as
\begin{eqnarray}\fl\Label{localeq}
n_{ij}^{(eq)}({\bf x},{\bf
p})&=&N_{i}(\b{x})\exp{\left(-(\b{p}-(\av{\b{p}_i}+\av{\b{p}_j})/2)^2/mk_B(T_i+T_j)\right)},
\end{eqnarray}
where  $T_i\equiv T_i(\b{x})$ and
$\av{\b{p}_i}\equiv\av{\b{p}_i(\b{x})}$ are average local
temperatures and momenta. Summing over the contributions from both
internal states gives
\begin{eqnarray}\Label{firstline}
\dot{n}_i({\bf x},{\bf p})|_{coll} &=& -\av{v_r(\b{x})}
_{ii}\sigma_{ii}(N_i({\bf x})/2)(n_i({\bf x}{\bf,
p})-n_{ii}^{(eq)}({\bf x},{\bf p}))\\\Label{secondline} &&-\langle
v_r(\b{x})\rangle _{ij}\sigma_{ij}N_j({\bf x})(n_i({\bf x},{\bf
p})-n_{ij}^{(eq)}({\bf x},{\bf p}))
\end{eqnarray}
where the factor of $1/2$ is necessary to prevent over counting
when determining the total collision rate between identical
particles. This form of damping conserves the collision invariants
and also allows a further simplification. From the definition of
the moments we have the identities
\begin{eqnarray}
N_1(\b{x})\av{\b{p}_1}+N_2(\b{x})\av{\b{p}_2}&=&N(\b{x})\av{\b{p}}\simeq 0,\\
N_1(\b{x})\av{\b{p}^2_1}+N_2(\b{x})\av{\b{p}^2_2}&=&N(\b{x})\av{\b{p}^2}.
\end{eqnarray}
so when the densities remain approximately similar it is then
reasonable to set $\av{\b{p}_1}+\av{\b{p}_2}=0$, and make the
replacement $T_1+T_2=2T$ in (\ref{localeq}). If we assume that
each momentum distribution is never far from its own local
equilibrium we may also neglect (\ref{firstline}), so that the
damping only arises from collisions between distinct states. This
leads to
\begin{eqnarray}\fl
&&\dot{m}({\bf x},{\bf p})|_{coll}=
\nonumber\\\fl&&-\av{v_r(\b{x})}_{12}\sigma_{12}\left(N_2(n_1(\b{x},\b{p})-n^{(eq)}_{12}(\b{x},\b{p}))-N_1(n_2(\b{x},\b{p})-n^{(eq)}_{21}(\b{x},\b{p})).\right)
\end{eqnarray}
which may be written as
\begin{eqnarray}\fl\Label{mrelaxation}
\dot{m}({\bf x},{\bf p})|_{coll}
&=&-\av{v_r(\b{x})}_{12}\sigma_{12}{N(\b{x})\over2}\left(m(\b{x},\b{p})-M(\b{x}){\exp\left(-\b{p}^2/2mk_BT\right)\over\sqrt{2\pi
mk_BT}}\right),
\end{eqnarray}
when the term proportional to $
n(\b{x},\b{p})-n^{(eq)}_{12}(\b{x},\b{p})-n^{(eq)}_{21}(\b{x},\b{p})$,
which is very small, is neglected. This means that the segregation
momentum distribution relaxes at approximately the same rate at
which each distribution would relax under the influence of
collisions with the other internal state.
 Reducing this to an effective one
dimensional rate requires that the distributions be radially
averaged. The result is simply the above form divided by the
circular transverse cross section $\pi
(2\sqrt{k_BT/m\omega_r^2})^2$. Using the thermal relative velocity, the collision term is now simply
\begin{eqnarray}\fl\label{relaxation}
\dot{m}(x,p)|_{coll}&=&-2a_{12}^2\omega_r^2\sqrt{m\over \pi
k_BT}N(x)\left(m(x,p)-M(x){\exp\left(-p^2/2mk_BT\right)\over\sqrt{2\pi
mk_BT}}\right).
\end{eqnarray}
The prefactor determines an effective segregation relaxation time
at the center of the trap of $\sim 30$ms for the experimental
parameters.

If there are coherences present between the two components the
above arguments are not complete because the coherence amplitude
provides an additional damping effect. It is not {\it a priori}
clear what sort of relaxation approximation should be applied to
the coherence, although the Cauchy-Schwartz inequality for the
system leads to
\begin{equation}
|F(\b{x})|^2 \leq N_1(\b{x})N_2(\b{x}),
\end{equation}
so that it appears reasonable to damp the coherence with the same
relaxation time approximation used for the ordinary distributions.
 In our simulations we
simply damp all $f_{lm}(\b{x},\b{p})$ at the same rate, but we modify the rate
to find the best agreement with the experiment, since at this
level of approximation the damping rate is effectively a free
parameter of the model. We find increasing the damping rate by a
factor of 2 gives reasonable agreement with the experiment.
Better agreement could be found by optimizing this correction for
each experimental run, but this not been pursued here.

\subsubsection{Effect of relative velocity} It is apparent from
the experiment that in some regions of the gas the relative
velocity of the two states becomes comparable to the thermal
velocity during segregation. We treat the effect of relative
velocity between the two distributions by including this into the
calculation of the collision time $\tau_d$. The average relative
velocity between particles in state $i$ with those in $j$ at
$\b{x}$ is
\begin{eqnarray}
\av{v_r(\b{x})}_{ij}=\int d^3\b{p}\int
d^3\b{p^\prime}{n_i(\b{x},\b{p})n_j(\b{x},\b{p^\prime})|\b{p}-\b{p^\prime}|\over
m N_i(\b{x})N_j(\b{x})}
\end{eqnarray}
Using $\b{Q}=\b{p}+\b{p^{\prime}}$, $\b{q}=\b{p}-\b{p^{\prime}}$;
assuming the momenta are distributed according to
\begin{eqnarray}
n_i(\b{x},\b{p})=N_i(\b{x})\exp{\left(-\alpha(\b{p}-\b{\bar{p}}_i(\b{x}))^2\right)}(\alpha/\pi)^{3/2}
\end{eqnarray}
where $\alpha=1/2mk_BT$, and choosing the mean local momenta to be
aligned with the $x$ axis
$\b{\bar{p}}_i(\b{x})=\bar{p}_i(x)\b{\hat{x}}$, the average is
\begin{eqnarray}\fl
\av{v_r(x)}_{ij}={(\alpha/\pi)^3\over
8m}\exp{\left(-\alpha(\bar{p_i}^2+\bar{p_j}^2)\right)}
\int d^3\b{Q}\exp{\left(-\alpha \b{Q}^2/2+\alpha\b{Q}\cdot(\b{\bar{p}}_i+\b{\bar{p}}_j)\right)}\nonumber\\
\times\int d^3\b{q}\;|\b{q}|\;\exp{\left(-\alpha
\b{q}^2/2+\alpha\b{q}\cdot(\b{\bar{p}}_i-\b{\bar{p}}_j)\right)},
\end{eqnarray}%---------------------------------------------------------------------------------
\begin{figure}
\begin{center}
\epsfig{file=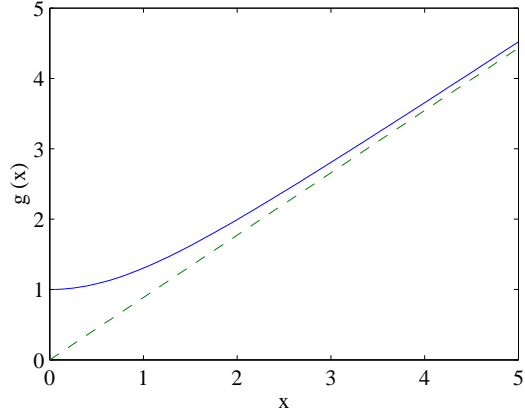,width=7cm} %\epsfxsize=\columnwidth
\end{center}
\caption{g$(x)$ and its asymptote.}\label{plotdamp}
\end{figure}
%---------------------------------------------------------------------------------
where the temperatures of the two components are treated as equal.
Since the integrals are spherically symmetric
\begin{eqnarray}\fl
\av{v_r(x)}_{ij}={(\alpha/\pi)^3\over
8m}\e^{-\alpha\left(\bar{p}_i^2+\bar{p}_j^2\right)}I_Q(\alpha/2,\alpha|\b{\bar{p}}_i
+\b{\bar{p}}_j|)I_Q(\alpha/2,\alpha|\b{\bar{p}}_i-\b{\bar{p}}_j|),
\end{eqnarray}
where $I_Q(a,b)$ and $I_q(a,b)$ are discussed in \ref{A2}. In
terms of the average thermal velocity (\ref{vtherm}) the final
result may then be written as
\begin{eqnarray}\Label{vdamp}
\av{v_r(x)}_{ij}=\bar{v}_{r}\;{\rm
g}\left({|\bar{\b{v}}_i-\bar{\b{v}}_j|\over\bar{v}_r\sqrt{\pi/4}}\right),
\end{eqnarray}
where
\begin{eqnarray}\Label{defineg}
{\rm
g}(x)={\e^{-x^2}\over2x}\left(x+(1+2x^2){\e^{x^2}\sqrt{\pi}\over2}{\rm
erf}(x)\right).
\end{eqnarray}
Since ${\rm g}(0)=1$ this reproduces the standard result for a
single component gas, and for the case where there is no mean
relative motion of the two internal states.  In the high relative
velocity limit obtained by setting
$\b{\bar{v}}_i(\b{x})=-\bar{\b{v}}_j(\b{x})=\b{\hat{x}}\bar{v}(x)$
with $ \bar{v}_r\ll \bar{v}(x)$, the large $x$ limit ${\rm
g}(x)\rightarrow \sqrt{\pi/4}x$ yields
\begin{eqnarray}
\av{v_r(x)}_{ij}\rightarrow 2\bar{v}(x).
\end{eqnarray}
Figure \ref{plotdamp} shows the simple interpolation given by g.
The effect of relative velocity on the relaxation may now be
included by substituting (\ref{vdamp}) for the thermal result used
in (\ref{mrelaxation}).
\subsubsection{Variation with temperature}
Since ${\rm g}(2)\simeq2$, when the axial
relative velocity is $\sqrt{\pi}\bar{v}_r\sim 1.7 \bar{v}_r$ the
damping rate will be doubled. For the JILA experiment the
modification is somewhat less because the thermal velocity is of
the order $\bar{v}_r\sim 20{\rm \;mm s^{-1}}$, whereas the
relative segregation velocity may be estimated to be at most $\sim
15{\rm \;mm s^{-1}}$, corresponding to a $50\%$ increase in
damping rate. This is expected to be important for the spin wave
experiment since coherence damping has significant effects on the
motion, and the coherence damping is also modelled using a simple
relaxation time approximation. Moreover, if the experiments are
attempted at higher temperatures this effect becomes more
significant because the timescale for initiation of segregation
scales like $\sim T^{-1/4}$ \cite{Williams2002}, and the cloud
length scales like $\sim T^{1/2}$ so that the segregation velocity
must behave like $\sim T^{3/4}$, whereas the thermal velocity
scales like $\sim T^{1/2}$, so the ratio of the segregation
velocity to the thermal velocity will be proportional to $\sim
T^{1/4}$. Clearly this correction will become relatively less
important for lower temperatures.

To find the effect of reducing the temperature on the coherence damping, we may compare the
temperature dependence of the rates (\ref{secondline}) and
(\ref{Boltzdamp}) with (\ref{cubicestimate}). In particular
(\ref{secondline}) and (\ref{Boltzdamp}) are proportional to
$\sqrt{k_BT}N(\b{x})$. The ratio of cubic to quadratic relaxation
rates will vary as $NT^{-3/2}$, so that this process is expected
to become more significant at lower temperatures. Since the peak
density also varies as $T^{-3/2}$ the ratio will vary as $T^{-3}$.
In particular, if the temperature is reduced by a factor of $3$
for the experimental parameters, the quadratic and cubic rates
(\ref{Boltzdamp}) and (\ref{cubicestimate}) will become similar.
The cubic collision terms (\ref{cubicf}) will be important in this
regime, and in the simplest approach, the relaxation time
approximation (\ref{secondline}) would need to be modified to
account for the relaxation caused by (\ref{cubicf}), with
$u_{ij}\simeq u_{12}$. One could also carry out moment equation
calculations similar to those of Nikuni \cite{Nikuni02} for the
cubic contribution to the damping in linearized spin equations,
but this not pursued here.

\section{Simulations}\label{sec:sim}
To simulate the experiment the equations
(\ref{neom},\ref{meom},\ref{feom}) are reduced to an effective one
dimensional description, and damping is included by adding the
simple relaxation approximation term (\ref{relaxation}) to the resulting
equations of motion.
\subsection{Simplifications}
 The high anisotropy of the trap and low densities
used allow the a reduction to a quasi-one dimensional model
starting from a noninteracting initial condition. We may also take
$\dot{N}({\bf x})=0$ as seen in the experiment, and verified by
our simulations.
\subsubsection{Initial motion}\label{initialmotion} The state of the gas before the
$\pi/2$ pulse, given by the
 Boltzmann distribution
\begin{eqnarray}\label{BoltzmannIC}
n({\bf x},{\bf p})|_{t=0}={\cal
N}\exp{\bigg(-{p^2\over2mkT}-{V_1({\bf x})+2u_{11}N({\bf x})\over
k_BT}\bigg)}
\end{eqnarray}
for the single internal state $|1\rangle$, is the stationary
solution of the equation of motion
\begin{eqnarray}\fl
\dot{n}({\bf x},{\bf p})=\left\{-({\bf p}/m)\cdot\nabla_{\bf x
}+\nabla(V_1({\bf x})+2u_{11} N({\bf x}))\cdot\nabla_{\bf
p}\right\}n({\bf x},{\bf p}).
\end{eqnarray}
 We may therefore neglect the density
dependent effective potential $2u_{11}N({\bf x})$ if
\begin{eqnarray}
\bigg|{2u_{11}\nabla N({\bf x})\over\nabla V_1({\bf
x})}\bigg|={2u_{11}N({\bf x})\over k_BT+2u_{11}N({\bf
x})}\leq{2u_{11}N({\bf x})\over k_BT}\ll 1.
\end{eqnarray}
For the densities used in our simulations $2u_{11}N(0)/kT\simeq
0.02$ allowing the use of the noninteracting initial condition.
After the pulse the system consists of an equal superposition of
two internal states in identical external configurations, so we
simply use
\begin{eqnarray}\label{nIC}
n({\bf x},{\bf p})|_{t=0}&=&{\cal
N}\exp{\bigg(-{p^2\over2mkT}-{V_1({\bf x})\over
k_BT}\bigg)},\\\label{mIC} m({\bf x},{\bf p})|_{t=0}&=&0,
\\\label{fIC} f({\bf x},{\bf p})|_{t=0}&=&{1\over 2}n({\bf x},{\bf
p})|_{t=0}.
\end{eqnarray}

\subsubsection{One dimensional equation}
The high aspect ratio of the trap ($33:1$) means that there is a
separation of timescales, leading to a physical transverse
averaging of the atomic motion. This means that we may average the
distributions over radial coordinates and momenta and find a one
dimensional equation of motion. The only change is that the
interaction strengths are divided by the transverse cross section
of the sample given by $\pi R_{\perp}^2$ where the transverse
radius $R_\perp$ is two standard deviations of the equilibrium density. A
typical interaction dependent term has the form
\begin{eqnarray}
{\partial f(\b{x},\b{p})\over\partial
t}=u\;G(\b{x})f(\b{x},\b{p}).
\end{eqnarray}
All of the terms can be seen to be either of this type, or to have
no interaction parameter and therefore to be unaltered by the
radial integration. Putting
\begin{eqnarray}
f(\b{x},\b{p})&=&f(x,p)\left({m\omega_r^2\over2\pi
k_BT}\right)\exp{\left(-m\omega_r^2(y^2+z^2)/2k_BT\right)}\nonumber\\
&&\times\left({1\over2\pi m k_BT}\right)\exp{\left(-(p_y^2+p_z^2)/2mk_BT\right)},\nonumber\\
G(\b{x})&=&G(x)\left({m\omega_r^2\over2\pi
k_BT}\right)\exp{\left(-m\omega_r^2(x^2+y^2)/2mk_BT\right)},
\end{eqnarray}
and integrating over radial momenta and coordinates leads to
\begin{eqnarray}
{\partial f(x,p)\over\partial t}&=&{u\over\pi(2\sqrt{\pi
k_BT/m\omega_r^2})^2}\;G(x)f(x,p)
\end{eqnarray}
where the denominator is clearly the transverse cross-section of
the sample, at a radius of two standard deviations. The effective
scattering parameter has the dimensions of energy$\times$length
required for a one dimensional interaction parameter.

\subsection{The differential potential}
The differential potential described by the parameter $\nu_{\rm
diff}$ is defined by the experimenters \cite{Lewandowski} as
\begin{eqnarray}\Label{nudiff}
\nu_{\rm diff}= {1\over 2\pi}\sqrt{{h\over
m}\left\langle{d^2\nu_{12}\over dz^2}\right\rangle},
\end{eqnarray}
where the Gaussian weighted average of the frequency shift
curvature is
\begin{eqnarray}\Label{gaussweight}
\left\langle{d^2\nu_{12}\over
dz^2}\right\rangle ={\int_{-3\bar{x}/2}^{3\bar{x}/2}dz\;N(z){d^2\nu_{12}(z)\over
dz^2}\over\int_{-3\bar{x}/2}^{3\bar{x}/2}dz\;N(z)},
\end{eqnarray}
and $\bar{x}$ is the half width at half maximum of the the axial
density distribution
\begin{eqnarray}
N(z)={1\over\sqrt{2\pi R^2}}\;\e^{-z^2/2R^2},
\end{eqnarray}
given by $\bar{x}=R\sqrt{2\log{2}}$, where
$R=\sqrt{k_BT/m\omega_z^2}$. It is apparent from Figure 2 of
\cite{Lewandowski} that we can accurately represent the gradient
of the differential potential in terms of a Gaussian with
amplitude $A_g$ defined as
\begin{eqnarray}\Label{gaussvdiff}
V_{\rm diff}(z)\equiv-A_gN(z).
\end{eqnarray}
We then obtain $A_g$ using (\ref{nudiff}),(\ref{gaussweight}) and
(\ref{gaussvdiff}), to find
\begin{eqnarray}
A_g&=&4\pi R^3m\omega_{\rm diff}^2\left\{{{\rm
erf}(3\bar{x}/2\sqrt{2}R)\over\sqrt{\pi}{\rm
erf}(3\bar{x}/2R)+(3\bar{x}/R)\exp{\left(-(3\bar{x}/2)^2/R^2\right)}}\right\}\nonumber\\
&\simeq&4\pi R^3m\omega_{\rm diff}^2\times0.44.
\end{eqnarray}
If we now expand (\ref{gaussvdiff}) for small $z$ we find
\begin{eqnarray}
-A_gN(z)\simeq {\rm constant} + {m\over2}\left(\omega_{\rm
diff}\sqrt{2}(2\pi)^{1/4}\sqrt{0.44}\right)^2z^2,
\end{eqnarray}
so that the true frequency near the center of the trap is $\sim
1.49\nu_{\rm diff}$. Such a correction is expected since the
Gaussian weighting process will reduce the resulting $\nu_{\rm
diff}$ by adding in negative frequencies, if the range of
integration is taken beyond the turning point of the Gaussian
$N(\b{x})$.

\subsection{Results}

%---------------------------------------------------------------------------------
\begin{figure}
\begin{center}
\epsfig{file=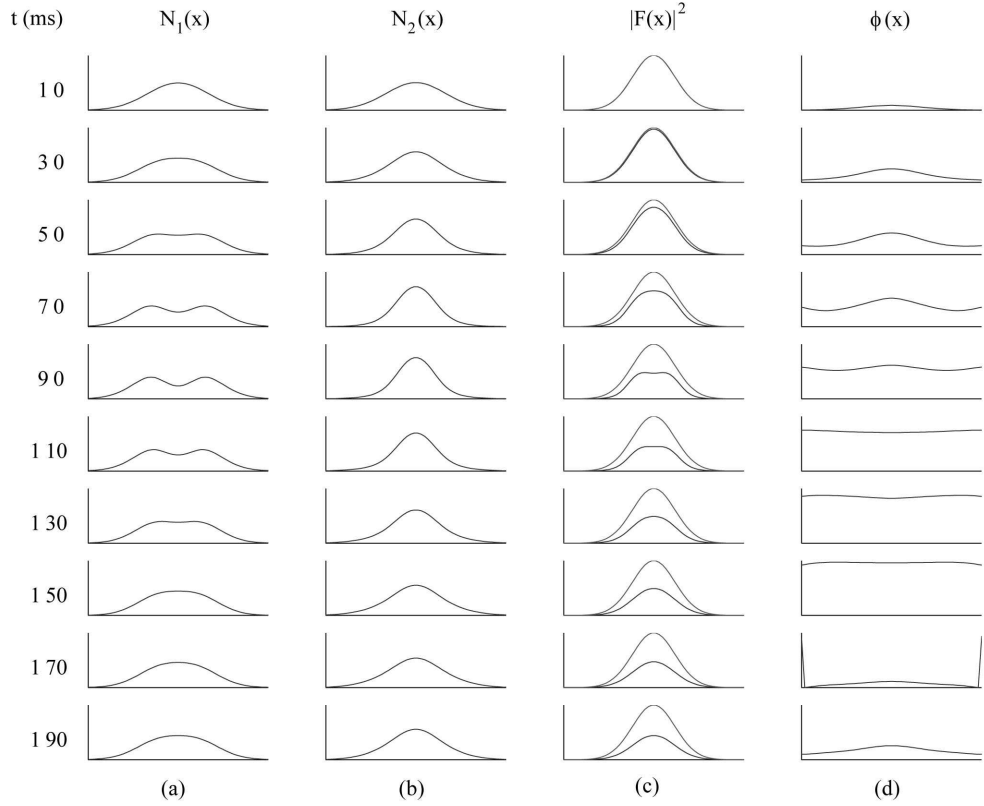,width=13cm}
\end{center}
\caption{Time is in ms down the left column. $N(0)=1.16\times
10^{13}{\rm cm}^{-3}$, $T=850$nK, $\nu_{\rm diff}=0.1$. (a) and
(b) are the densities of the two internal states in arbitrary
units. (c) Shows the coherence density, with initial condition for
comparison, and d) shows the relative phase between the two
internal states $\phi(x)={\rm angle}(F(x))$; the plot range is
$[0,2\pi)$.}\label{fig1}
\end{figure}
%---------------------------------------------------------------------------------

%---------------------------------------------------------------------------------
\begin{figure}
\begin{center}
\epsfig{file=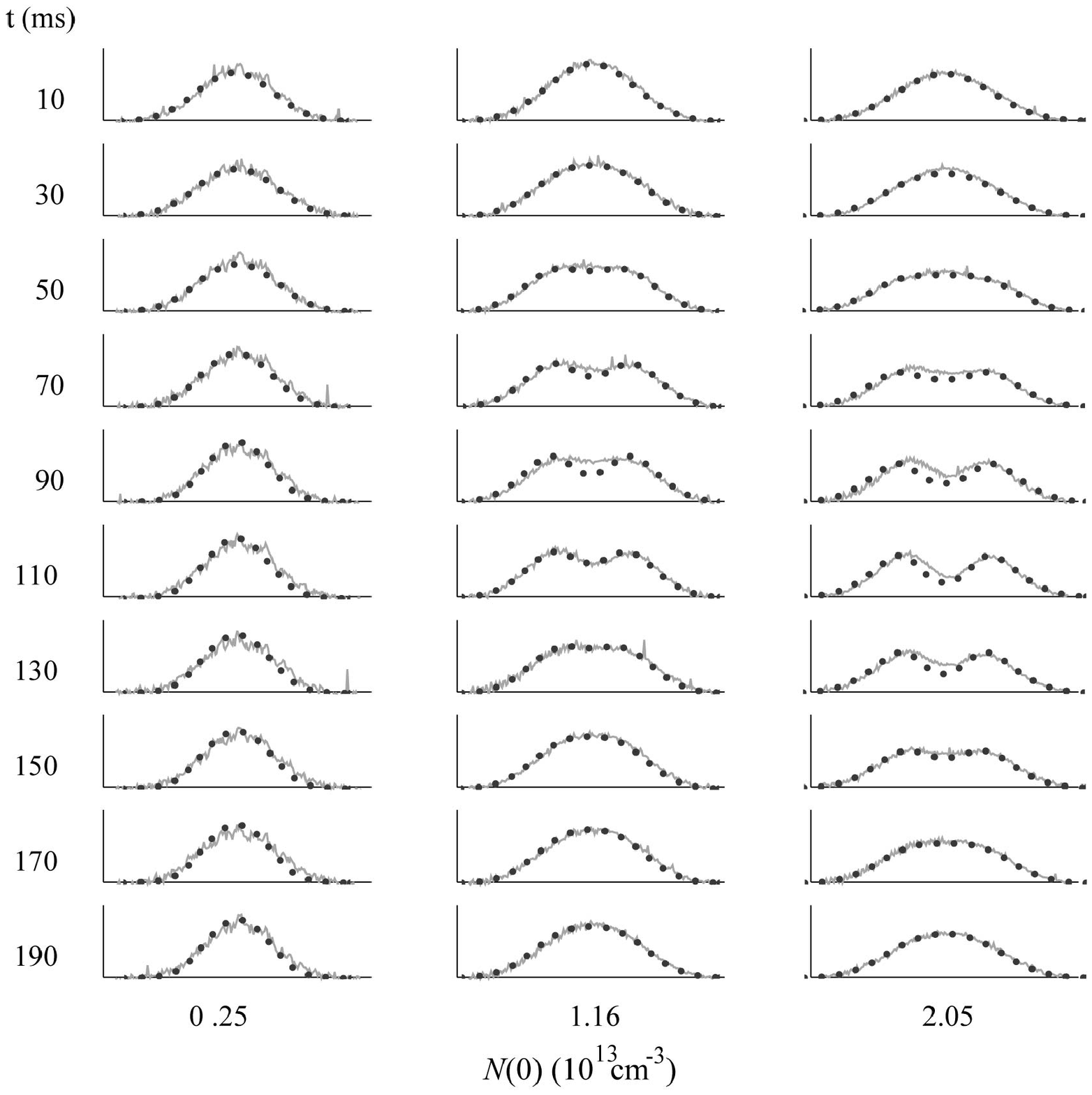,width=13cm} \epsfxsize=\columnwidth
\end{center}
\caption{Measured $N_1(z)$, with simulations (dotted line) for
$\nu_{\rm diff}=0.1$ and peak densities listed at the foot of each
column. Time is in ms down the left column.} \label{fig2}
\end{figure}
%---------------------------------------------------------------------------------
Figure \ref{fig1} shows the result of simulating equations
(\ref{neom},\ref{meom},\ref{feom}) with $\nu_{\rm
 diff}=0.16\;{\rm Hz}, N(0)=1.8\times10^{13}{\rm cm}^{-3}$, with the relaxation
 time approximation (\ref{relaxation}) modelling the damping.
 The effect is very similar to that seen in the experiment \cite{Lewandowski}. In
(a) and (b) the axial density profile of the $\ket{1}$ and
$\ket{2}$ states show a clear axial segregation of the two
internal states. (c) Shows the decay of the coherence which causes
the transience of the phenomena on the timescale of the
experiment. The initial condition is shown for comparison. Column
(d) is the phase of $F(\b{x})$, from which it is clear that the
relative motion tends to smooth out the relative phase gradient
that is caused by the differential potential $V_{\rm
diff}(\b{x})$.

Figure \ref{fig2} shows the results of simulations for the
experimental parameters, with the measured data \cite{citedata}.
The timescale for decay of coherence is well described by the
simple relaxation time approximation, with an extra factor of 2.
The initiation times, relaxation times and amplitude of the
segregation agree qualitatively with the experiment.

\begin{figure}
\begin{center}
\epsfig{file=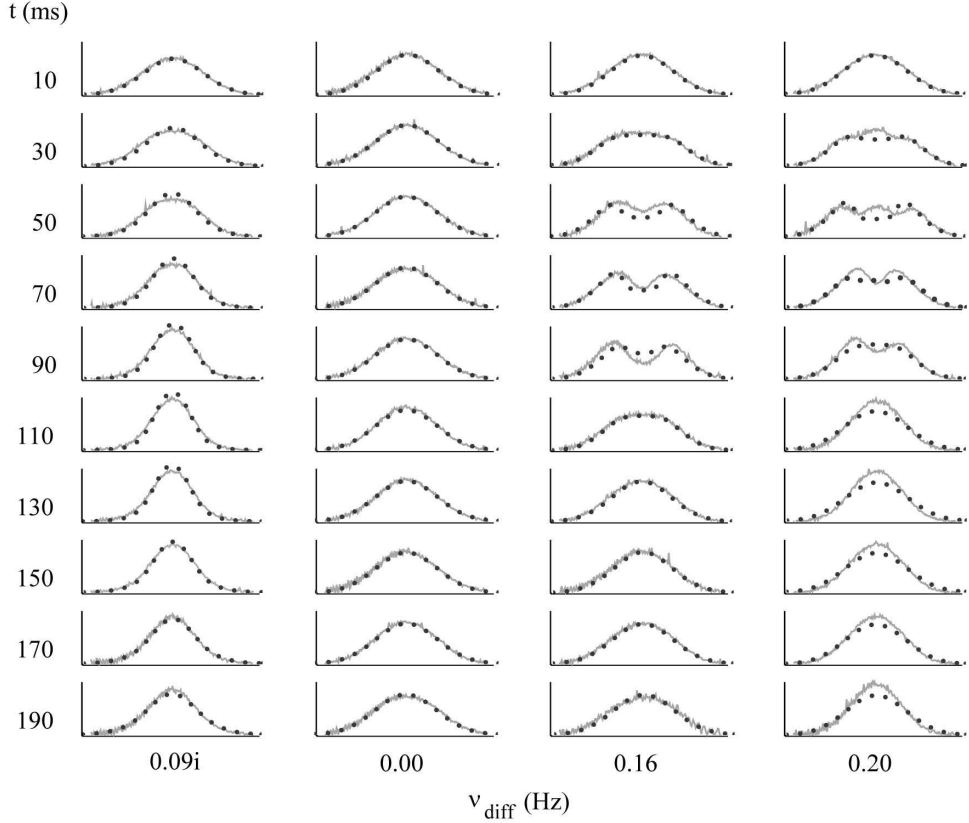,width=13cm} \epsfxsize=\columnwidth
\end{center}
\caption{Measured $N_1(z)$, with simulations (dotted line) for
peak density $N(0)=1.8\times10^{13}\;{\rm cm}^{-3}$ and $\nu_{\rm
diff}$ listed at the foot of each column. Time is in ms down the
left column.} \label{fig3}
\end{figure}

Figure \ref{fig3} shows the variation with $\nu_{\rm diff}$, and
the numerical results. We see here that what is described as
"higher order effects" are not well modelled by the simulation for
$\nu_{\rm diff}=0.20\;{\rm Hz}$. The initiation time and
transience are still in good agreement but the extra peak in the
center of $N_1(x)$ does not emerge.

\subsection{`Classical' Motion}
It is useful to consider the classical motion that would arise if
the system could be prepared with identical initial state
distributions as constructed by the first $\pi/2$ pulse of the
experiment, while also setting the coherence to zero. If coherence
is neglected the initial condition (\ref{BoltzmannIC}) is merely a
density perturbation from the new Boltzmann equilibrium
\begin{eqnarray}
n({\bf x},{\bf p})&=&{\cal N}\exp{\left(-{p^2\over2mk_BT}-{V_{\rm
eff}({\bf x})\over k_BT}\right)}\cosh{\left({V_{\rm diff}({\bf
x})\over k_BT}\right)},
\nonumber\\
m({\bf x},{\bf p})&=&{\cal M}\exp{\left(-{p^2\over2mk_BT}-{V_{\rm
eff}({\bf x})\over k_BT}\right)}\sinh{\left({V_{\rm diff}({\bf
x})\over k_BT}\right)}.
\end{eqnarray}
 The corresponding single species distributions are
\begin{eqnarray}
n_1({\bf x},{\bf p})&=&{\cal
N}_1\exp{\left(-{p^2\over2mk_BT}-{V_1({\bf x})+2u_{11}N_1({\bf
x})+u_{12}N_2({\bf x})\over k_BT}\right)},\nonumber\\
n_2({\bf x},{\bf p})&=&{\cal
N}_2\exp{\left(-{p^2\over2mk_BT}-{V_2({\bf x})+2u_{22}N_2({\bf
x})+u_{12}N_1({\bf x})\over k_BT}\right)}\nonumber\\
\end{eqnarray}
for appropriate normalizations ${\cal N}_1$ and ${\cal N}_2$. We
have seen in section \ref{initialmotion} that the interactions
have a negligible effect on the initial condition, and similar
reasoning shows the equilibrium segregation after the pulse is
negligible for the densities used in the experiment
\cite{Lewandowski}. It follows that the classical motion towards
the new equilibrium would be small, as long as the fluctuations
caused by the perturbation are also small.
%---------------------------------------------------------------------------------
%---------------------------------------------------------------------------------
\section{Spin waves and atomic motion}\label{sec:spin}
In the experimental report it is asserted that the motion observed
must be due to actual physical motion of the atoms since energy
conservation would seem to prohibit spontaneous interconversion of
the internal states \cite{Lewandowski}. In this situation `motion'
refers to the redistribution of the atomic density in the trap.
The issue is whether or not it is possible to distinguish between
two interpretations: The apparent motion is the result of either
\begin{itemize}
\item[i)] Transport of the atoms in the relative phase
gradient arising from the Zeeman and mean field effects, or
\item[ii)]The coherent interconversion of the internal
states as a result of the interactions.
\end{itemize}
In defence of the first picture, it has been suggested in
\cite{Lewandowski} that the rate of segregation can be accounted
for if the small differential potential is somehow amplified in a
coherent collision processes which channels the thermal energy of
atoms in a particular direction. In the second picture, the
collisions between the two different internal states are
interpreted as causing a rotation in pseudo-spin space of the
spinor wave function for the two state superposition of the field
operators.

It is important to note here that if the transport terms are
neglected in the equations of motion, segregation does {\it not}
occur. The equations of motion for the densities are found by
integrating the equations of motion (\ref{neom}), (\ref{meom}),
(\ref{feom}) over momentum which leads to
\begin{eqnarray}
\dot{N}({\bf x})+\nabla\cdot(N({\bf x}){\bf v}_{N}({\bf x}))&=&0\\
 \dot{M}({\bf x})+\nabla\cdot(M({\bf x}){\bf v}_{M}({\bf x}))&=&0\\
 \dot{F}({\bf x})+\nabla\cdot(F({\bf x}){\bf
v}_{F}({\bf x}))&=&-i\omega_{R}({\bf x})F({\bf x}),
\end{eqnarray}
where the velocities are, for example
\begin{eqnarray}
{\bf v}_{N}={1\over m N({\bf x})}\int d^3{\bf p}\;{\bf p}\;n({\bf
x},{\bf p}),
\end{eqnarray}
and the Ramsey frequency is
$\omega_R(\b{x})=\Delta\omega_c(\b{x})+\Delta\omega_z(\b{x})$.

If the transport terms are neglected the velocities vanish, and no
change in the densities can occur---the only effect is to make a
redistribution in momentum space.  Inclusion of the transport terms
then translates this redistribution in momentum space into a
redistribution in position.  In this sense, we agree with the view of
\cite{Lewandowski} that the segregation is caused by actual motion of
the atoms.  The effect is however quite large, since it involves more
forces than simply the classical forces induced by the gradient of the
differential potential, namely the forces caused by the production of a
spatially dependent phase of $f({\bf x},{\bf p})$, as seen in the
second line of (\ref{feom}).

%It would thus seem that the segregation effect is the result of
%{\it relative motion} of the two states in a {\it relative phase
%gradient}, which is enhanced by the velocity changing collisions,
%as foreseen in \cite{Lewandowski}.

\section{Comparison with other work}\label{sec:comp}
While preparing this work several other papers on the JILA
experiment \cite{Lewandowski} have appeared
\cite{Oktel,Fuchs,Williams2002}. The works
\cite{Williams2002,Fuchs} use essentially the same formalism as
our own, with slightly different degrees of approximation. In
\cite{Williams2002} the simulations agree with experiment to the
same quantitative degree as our own, although they obtain better
results for the variation of $\nu_{\rm diff}$, in particular we do
not see higher order effects for the highest value $\nu_{\rm
diff}=0.2\;{\rm Hz}$ as reported in the experiment and found by
Williams {\it et al} \cite{Williams2002}. In \cite{Fuchs} the
simulated initiation times appear more rapid than seen in the
experiment. Our results show better agreement for this feature,
which may be because we have inverted the curvature averaging
process used to characterize the differential potential. In
\cite{Williams2002,Fuchs} the timescale of initiation for the
segregation is found analytically, and in \cite{Williams2002} it
is noted that this also gives the position of the nodes of
$M(\b{x})$. In \cite{Oktel} a so called Landau-Lifshitz equation
of motion was obtained, and solved numerically, finding
qualitatively similar behavior, although the timescale of
evolution is significantly longer than that observed in the
experiment.

More recent work at JILA \cite{McGuirk02} has focused on providing
data for a collective mode analysis which has been carried out by
Nikuni {\it et al} \cite{Nikuni02} using a truncated moment
equation approach. In this experiment the differential potential
is controllable in real time, so that a small amplitude
fluctuation may be excited, and the potential gradient then set to
zero for the remainder of the motion. The same relaxation time
approximation is used to that derived here, and the simulations of
\cite{Nikuni02} show excellent agreement with the data of
\cite{McGuirk02}. Some discrepancy is found for the quadrupole
mode in the regime where Landau damping is the dominant form of
dissipation, and the authors note that the details of the trap are
likely to be important in describing this process accurately.
\section{Conclusions}\label{sec:conc}
We have simulated the JILA experiment using the Wigner function
approach and found reasonable agreement with the data of
\cite{Lewandowski}. We have discussed the Ramsey frequency and the
local transition frequency for this system, and emphasized the
difference of the two, finding that the Ramsey technique is
insensitive to the decay of coherence. A relaxation time
approximation for the experiment has been derived, including the
effect of relative velocity which is expected to be important for
higher temperature experiments. A Quantum Boltzmann collision term
for two species with arbitrary S-wave interaction strengths was
found, and we evaluated the effects of scattering on the coherence
for non-condensed \rubidium\; using a Boltzmann equilibrium form
for the momentum distributions. It is apparent that the cubic
terms in the collision integral depend on the phase space density
and will be important for more degenerate regimes.
\appendix
\section{Continuum limit}\label{A1}
The derivation of (\ref{collisionintegral}) from the terms
(\ref{directav}) and (\ref{exchangeav}) is simplified by the
identities of the form
\begin{eqnarray}\fl
\sum_{ijIJ,\b{e}}\delta(\Delta\omega(\b{e}))\av{U_{IJ}^{\dagger}(\b{e})[\psi^{\dagger}_{l\b{k}}\psi_{m\b{k}},U_{ij}(\b{e})]}^{*}
=\sum_{ijIJ,\b{e}}\delta(\Delta\omega(\b{e}))\av{[U_{ij}(\b{e}),\psi^{\dagger}_{m\b{k}}\psi_{l\b{k}}]U_{IJ}^{\dagger}(\b{e})}\nonumber\\\fl
\end{eqnarray}
so that (\ref{directav}) and (\ref{exchangeav}) are complex
conjugates with $l \leftrightarrow m$. The Quantum Boltzmann
equation (\ref{collisionintegral}) is then obtained by using
\begin{eqnarray}
f_{lm}(\b{k})&=&\left({\pi\over\Delta}\right)^3\sum_{\b{N}}{\rm
Tr}\{v_{\b{N}}\psi^{\dagger}_{l\b{k}}\psi_{m\b{k}}\}\nonumber\\
&=&\left({\pi\over\Delta}\right)^3\av{\psi^{\dagger}_{l\b{k}}\psi_{m\b{k}}},
\end{eqnarray}
 so that the commutators may be evaluated and we may carry out the procedures in
\cite{QKI} for passing to the continuum limit. This involves
factorizing the operator averages and using the local equilibrium
forms
\begin{eqnarray}\fl
\av{\psi^{\dagger}_{l\b{K}_1}(\b{x})\psi_{m\b{K}_2}(\b{x^{\prime}})}&=&{\rm
g}(\b{x}-\b{x^{\prime}})\;f_{lm}\left({\b{x}+\b{x^{\prime}}\over2},\b{K}_1\right)\delta_{\b{K}_1,\b{K}_2}\\\fl
\av{\psi_{l\b{K}_1}(\b{x})\psi^{\dagger}_{m\b{K}_2}(\b{x^{\prime}})}&=&{\rm
g}(\b{x}-\b{x^{\prime}})\left(f_{ml}\left({\b{x}+\b{x^{\prime}}\over2},\b{K}_1\right)+\delta_{lm}\right)\delta_{\b{K}_1,\b{K}_2},
\end{eqnarray}
and the continuum limit of the summation
\begin{eqnarray}\Label{continuum1}
\sum_{\b{K}_2,\b{K}_3,\b{K}_4}&&\left({\pi\over\Delta}\right)^3M_\Delta(\b{K}+\b{K}_2-\b{K}_3-\b{K}_4)\nonumber\\
&&\rightarrow{1\over(2\pi)^6}\int\;d^3\b{K}_2\int\;d^3\b{K}_3\int\;d^3\b{K}_4\nonumber\\
&&\times\delta(\b{K}+\b{K}_2-\b{K}_3-\b{K}_4),
\end{eqnarray}
where
\begin{eqnarray}
{\rm g}(\b{x})={1\over\pi^3}\left[{\sin{\Delta x}\over
x}\right]\left[{\sin{\Delta y}\over y}\right]\left[{\sin{\Delta
z}\over z}\right],
\end{eqnarray}
and $M_\Delta(\b{Q})$ is the approximate delta function arising
from the factorization of operator averages
\begin{eqnarray}\Label{continuum2}
M_\Delta(\b{Q})&=&\int\;d^3\b{y}\int\;d^3\b{y^{\prime}}{\rm
g}(\b{y}){\rm g}(\b{y^{\prime}})[{\rm
g}(\b{y}-\b{y^{\prime}})]^3\e^{i\b{Q}\cdot(\b{y}-\b{y^{\prime}})}\nonumber\\
&=&\prod_{i=1}^3\left({\Delta\over\pi}\right)^3\left\{{2\over3}\delta_{Q_i,0}+{1\over3}\delta_{Q_i,\Delta}+{1\over3}\delta_{Q_i,-\Delta}\right\}.
\end{eqnarray}
(This corrects an error in Eqs. (128) and (129) of \cite{QKI},
which led to an extra factor of $(2\pi)^3$ in the final
Uehling-Uhlenbeck collision term (132)).

\section{Integrals}\label{A2}
Some useful integrals for this paper are
\begin{eqnarray}
I_Q(a,b)&=&\int_0^{2\pi} d\phi\int_0^\pi\; d\theta\;\sin\theta
\int_0^\infty dQ\; Q^2\e^{-aQ^2+bQ\cos\theta}\\
&=&{2\pi\sqrt{\pi}\over
ab}\left({b\over2\sqrt{a}}\right)\exp{(b^2/4a)}
\end{eqnarray}
and
\begin{eqnarray}
I_q(a,b)&=&\int_0^{2\pi} d\phi\int_0^\pi\;d\theta\;\sin\theta
\int_0^\infty dq\;q^3\e^{-aq^2+bq\cos\theta}\\
&=&{2\pi\e^{b^2/4a}\over a^2}\;{\rm g}\left(b/2\sqrt{a}\right),
\end{eqnarray}
where
\begin{eqnarray}
{\rm
g}(x)={\e^{-x^2}\over2x}\left(x+(1+2x^2){\e^{x^2}\sqrt{\pi}\over2}{\rm
erf}(x)\right).
\end{eqnarray}
Defining
\begin{eqnarray}
B(\b{p})\equiv
(2\pi\hbar)^3\left({\alpha\over\pi}\right)^{3/2}\exp{(-\alpha
\b{p}^2)}
\end{eqnarray}
 the integral
\begin{eqnarray}\fl
I_{34}&\equiv&\int\;d^3\b{P}_1\int\;{d^3\b{P}_2\over(2\pi\hbar)^3}\int\;{d^3\b{P}_3\over(2\pi\hbar)^3}\int\;{d^3\b{P}_4\over(2\pi\hbar)^3}B(\b{P}_3)B(\b{P}_4)
\nonumber\\\fl
&&\;\;\;\;\;\;\;\;\;\;\;\;\;\;\times\delta(\b{P}_1+\b{P}_2-\b{P}_3-\b{P}_4)\delta(\epsilon_1+\epsilon_2-\epsilon_3-\epsilon_4)
\end{eqnarray}
is found by using the transformation $\b{Q}=\b{P}_1+\b{P}_2$,
$\b{q}=\b{P}_1-\b{P}_2$ to obtain
\begin{eqnarray}\fl
I_{34}&=&{m\over2^4(2\pi\hbar)^3}\left({\alpha\over\pi
}\right)^3\int\;d^3\b{Q}\int\;d^3\b{K}\int\;d^3\b{q}\int\;d^3\b{k}\;\exp{(-\alpha(\b{Q}^2+\b{q}^2)/2)}
\nonumber\\\fl
&&\;\;\;\;\;\;\;\;\;\;\;\;\;\;\times\delta(\b{q}^2-\b{k}^2)\delta(\b{Q}-\b{K})\nonumber\\\fl
&=&{2\pi m\over2^4(2\pi\hbar)^3}\left({\alpha\over\pi
}\right)^3I_Q(\alpha/2,0)I_q(\alpha/2,0)=\sqrt{2\pi\over\alpha}{2m\over(2\pi\hbar)^3}.
\end{eqnarray}
The integral $I_{234}$ is found in a similar fashion as
\begin{eqnarray}
{\rm
I}_{234}&\equiv&\int\;{d^3\b{P}_1\over(2\pi\hbar)^3}\int\;{d^3\b{P}_2\over(2\pi\hbar)^3}\int\;{d^3\b{P}_3\over(2\pi\hbar)^3}\int\;d^3\b{P}_4\;B(\b{P}_2)B(\b{P}_3)B(\b{P}_4)
\nonumber\\\fl
&&\;\;\;\;\;\;\;\;\;\;\;\;\;\;\times\delta(\b{P}_1+\b{P}_2-\b{P}_3-\b{P}_4)\delta(\epsilon_1+\epsilon_2-\epsilon_3-\epsilon_4)\nonumber\\\fl
&=&{\sqrt{3}\over4\pi k_BT}.
\end{eqnarray}
\section*{Acknowledgements}
We would like to thank Rob Ballagh for his numerical expertise and
Eric Cornell for informing us about the experiment
\cite{Lewandowski}. Funding for this project was provided by the
Marsden Fund under contract PVT-902.
\bibstyle{plain}

\end{document}